\documentclass[12pt]{article}

\usepackage[round,comma,numbers,sort&compress]{natbib}
\usepackage{times}
\usepackage{url}
\usepackage[table]{xcolor}
\usepackage{amsmath}
\usepackage{amsfonts}
\usepackage{hyperref}
\usepackage{booktabs}
\usepackage{graphicx}
\graphicspath{{figs/}}

\usepackage{dcolumn} 
\newcolumntype{d}[1]{D{.}{.}{#1}}

\usepackage[table-number-alignment=center, group-separator={}]{siunitx}
\def\sym#1{\ifmmode^{#1}\else\(^{#1}\)\fi}
\usepackage{makecell}

\newcommand{\fref}[1]{Fig.~\ref{#1}}
\newcommand{\tref}[1]{Table~\ref{#1}}

\newcommand{\sm}[1]{\textcolor{blue}{\textit{SI} #1}}

\title{Author Mentions in Science News Reveal Widespread Disparities Across Name-inferred Ethnicities}

\author
{Hao Peng$^{1,2,3*}$, Misha Teplitskiy$^{1,4}$, David Jurgens$^{1,5*}$\\
\normalsize{$^{1}$School of Information, University of Michigan, Ann Arbor, MI, USA}\\
\normalsize{$^{2}$Kellogg School of Management, Northwestern University, Evanston, IL, USA}\\
\normalsize{$^{3}$Northwestern Institute on Complex Systems, Northwestern University, Evanston, IL, USA}\\
\normalsize{$^{4}$Laboratory for Innovation Science at Harvard, Harvard University, Boston, MA, USA}\\
\normalsize{$^{5}$Department of Computer Science \& Engineering, University of Michigan, Ann Arbor, MI, USA}\\
\normalsize{$^{*}$To whom correspondence may be addressed.}\\
\normalsize{Email: jurgens@umich.edu or hao.peng@kellogg.northwestern.edu}
}
\date{}

\begin{document}

\baselineskip20pt
\maketitle

\begin{abstract} 

Media outlets play a key role in spreading scientific knowledge to the general public and raising the profile of researchers among their peers. 
Yet, how journalists choose to present researchers in their stories is poorly understood. 
Using a comprehensive dataset of 223,587 news stories from 288 U.S. outlets reporting on 100,486 research papers across all areas of science, we investigate if the authors' ethnicities, as inferred from names, are associated with whether journalists explicitly mention them by name.
By focusing on research papers news outlets chose to cover, our analysis reduces concerns that differences in name mentions are driven by differences in research quality or newsworthiness.
We find substantial disparities in name mention rates across ethnically-distinctive names.
Researchers with non-Anglo names, especially those with East Asian and African names, are significantly less likely to be mentioned in news stories covering their research, even when comparing stories from a particular news outlet reporting on publications in a particular scientific venue on a particular research topic. 
The disparities are not fully explained by authors' affiliation locations, suggesting that pragmatic factors such as difficulties in scheduling interviews play only a partial role.
Furthermore, among U.S.-based authors, journalists more often use authors' institutions instead of names when referring to non-Anglo-named authors, suggesting that journalists' rhetorical choices are also key.
Overall, this study finds evidence of ethnic disparities in how researchers are described in the media coverage of their research, likely affecting thousands of non-Anglo-named scholars in our data alone.

\end{abstract} 

\clearpage

\section{Introduction}

Scientific breakthroughs often attract media attention, which serves as a key mechanism for public dissemination of new knowledge \cite{scheufele2013communicating, brossard2013science}. Science media coverage not only distills research insights but also puts a face on who was responsible for the research. The media coverage can then feed back into researchers' careers \cite{fanelli2013any}. 
Furthermore, science news reporting may over time shift the public's perception of \textit{who} a scientist is \cite{miller2018development}. 
Under-representing particular demographic groups can perpetuate the view that scientists are white men \cite{turner2008faculty, banchefsky2016but}, and potentially weaken the pipeline of recruiting diverse students into academic careers \cite{reuben2014stereotypes, hill2018discovery, hofstra2020diversity}.

Science media coverage can be separated into two stages: (i) the likelihood of coverage---whose paper gets reported and (ii) the \textit{quality of coverage}---i.e., given that a paper is reported, how is the depth of coverage? For example, high quality coverage may describe the sophistication of the research, eminence of the scholars, or credit them in quotes and name attributions. 

Recent research has examined disparities in the first stage, by comparing the coverage distribution to the overall distribution of published scholarship, i.e., P(paper reported $|$ paper published) \cite{vasarhelyi2021gender, chapman2022altmetric, davidson2023analysis}.
These studies find disparities in the amount of media coverage by author gender and country. 
Yet coverage can differ not only in likelihood, but also in its depth and  quality. The contribution of this paper is to examine ethnicity-related disparities in coverage quality among scientific papers receiving at least some coverage. There are substantial trade-offs in this choice of research design. On the one hand, focusing on the set of papers already deemed newsworthy sidesteps some of the potential confounds affecting the association between ethnicity and coverage, in particular research topic, newsworthiness, and quality. On the other hand, the design leaves outside its purview the many other ways in which scientists' ethnicity may affect media coverage, including potential biases in the selection of individuals into topics \cite{kozlowski2022intersectional}, institutions, publishing outlets \cite{peng2021acceptance}, coverage amount, and so on. It is possible that many of these disparities compound across stages, making those found in any one stage a severe underestimate of the overall extent. We highlight that this paper contributes only a piece to a much larger puzzle.

Here, we focus on one highly salient feature related to the quality of coverage---whether the author is mentioned or credited by name in the stories covering their research, i.e., P(name mentioned $|$ paper reported).
By focusing on \textit{how} rather than \textit{whether} reporters choose to report a scientific paper, our research design can precisely control for papers' newsworthiness and analyze important factors related to author demographics.


In writing about specific scientific advances, journalists face choices over how much attention to devote to each relevant researcher, and whom to ignore altogether. 
Empirical and theoretical literature motivates the possibility that ethnic disparities exist in journalists' choices of whom to feature and the nature of the resulting coverage \cite{callison2019reckoning, robinson2019white, sui2018role}. 

Empirically, a number of studies have established gender and ethnic disparities in conventional scientific outcomes, such as funding \cite{ginther2011race, oliveira2019comparison, hoppe2019topic}, publications \cite{way2016gender, peng2021acceptance}, and citations \cite{lariviere2013bibliometrics, huang2019historical}, as well as scientists' online visibility \cite{vasarhelyi2021gender, peng2022gender}. 
Furthermore, research points to demographic disparities and the stereotyped media coverage of the general population \cite{behm2013race, jia2015measuring, jia2016women, merullo2019investigating, smith1997all, devitt2002framing}.
The presence of abundant ethnic bias in traditional media suggests that the disparities may appear in how science is covered at the very latter stage as research disseminates to the public. 

Theoretically, we hypothesize a number of mechanisms that may produce ethnic disparities in which authors are mentioned in media coverage of science, and test them where possible. 
First, U.S.-based journalists may face pragmatic difficulties in interviewing researchers in distant time-zones and possibly with country-specific adoptions of different communication technologies \cite{lin2022remote}. 
Second, even for authors located within the same geographical region (e.g., in the U.S.), certain authors may have limited proficiency in speaking English. 
Furthermore, journalists may rely on their professional networks to contact sources. 
Analyses of the media landscape in the U.S. \cite{grieco2018newsroom, clark2018asne} and other markets \cite{nielsen2020race} show that the demographics of journalists and editors are highly unrepresentative of the broader populations. The demographics of journalists are likely to correlate with that of individuals in their professional networks \cite{mcpherson2001birds}, suggesting that the researchers journalists can reach most readily are also unrepresentative. To the extent that these pragmatic factors---interviewing difficulties and professional networks---correlate with the perceived ethnicities of names, certain researchers may be more or less mentioned.

Third, while science journalists aim to write stories that appear credible to their audiences \cite{sundar1998effect}, they may lack direct information on the credibility of authors of the relevant research papers and may not have the time to acquire such information. Facing unfamiliar names and time constraints, journalists may rely on stereotypes, inferring for example that some researchers are less competent or authoritative on some topics than others, or expecting their audiences to harbor such perceptions. Prior research has found such stereotyping in the context of researcher gender and gender-typical research topics \cite{knobloch2013matilda}. Inferences of competence and authoritativeness can lead journalists to choose some names over others, which is a form of statistical discrimination \cite{lang2012racial, neumark2018experimental}.
Fourth, journalists may not be the relevant actors at all. Some news coverage originates from press releases created by in-house public relation staff at universities. News outlets often reprint these press releases in part or in full, and any disparities therein may thus be passed on directly to the outlets and their audiences.

Here, we present the first large-scale and science-wide analysis of ethnic disparities in author mentions in science news covering research papers and explore the mechanisms producing them based on a computational analysis of 223,587 news stories mentioning 100,486 published papers.
\textit{By focusing on papers that already were deemed newsworthy, our research design side-steps the question of whose research is covered in the news in the first place, choices which may themselves be associated with ethnicity.}

We use the term ``ethnicity'' rather than nationality or race for two reasons: (1) an author's nationality is largely masked by their affiliations and is very fluid, especially in the U.S. context; 
(2) journalists only have access to author names upon reading the paper, and name reflects cultural origin that is more related to ethnicity and signals a richer set of information than race.

Lacking the information about authors' self-identity, we based our study on the \textit{perceived} ethnicity inferred from names to distinguish it from authors' true ethnicity.
This research choice entails substantial trade-offs. In fact, authors' self-identities may differ from their perceived ones, and some authors may self-identify with more than one ethnicity. In some cases, journalists know authors' self-identified ethnicity. Nevertheless, in many cases, journalists will not know how authors self-identity themselves and rather infer them from names \cite{crabtree2022racially, crabtree2023validated}. In these cases, using authors' self-identities would be problematic, as it would misrepresent the actual perceptions journalists form and possibly use when they write their stories. 

Our adoption of the perceived-ethnicity construct and the operationalization of it via names have three merits: (1) using the perception of ethnicity enables us to measure disparities in the imperfect information environment journalists actually face where the ethnicity of the author is not known upon first seeing the paper, and is thus more likely to illuminate their decision processes; (2) the construct of perceived ethnicity inferred via names has been widely used for decades in audit studies that use names to signal ethnicity or race to evaluators \cite{bertrand2004emily, gaddis2015discrimination, gaddis2020searching, gaddis2017black} and shown to be a highly effective proxy for studying disparities at scale; (3) the literature suggests that self-identified ethnicity and the perceived one are highly correlated, and that humans can and do infer ethnicity from names fairly accurately \cite{gaddis2017black, gaddis2017racial, sood2018predicting, sweeney2013discrimination}. 

\section{Data and Analysis Design}

We constructed a multidisciplinary dataset by combining news stories with metadata of the scientific papers they cover, and then infer demographic attributes of the papers' authors based on their names.
Our final corpus of news-reported papers was sourced from \textit{Altmetric.com}, which consists of 223,587 news stories from 288 U.S.-based outlets reporting on 100,486 scientific papers, with a total of 276,202 story-paper mention pairs. 
For all these new stories, we have the textual content, the mention date, and the outlet information.

We adopted a regression model to examine whether the paper's authors are mentioned by name in the news story.
Because journalists can mention and feature several authors when covering a paper in their news story, we treated each (story, paper, author) triplet as an observation in the regression.
Since the first author and the last author often contribute most to the work and are recognized as such in science journalism guidelines \cite{blum2006field}, we included them in our analysis by default. We also included any additional corresponding author of a paper. 
Thus, for each paper, we focused on authors at the highest ``risk'' of being mentioned in the story including first author, last author, and all authors designated as the corresponding author.
As a result, there are 524,052 observations in our regression analyses (\sm{Detailed Dataset Description}).

We additionally obtained papers' metadata from the Microsoft Academic Graph and the Web of Science databases \cite{wang2019review, peng2021neural}, including author name, author rank, corresponding author status, authorship position, affiliation rank and location, number of authors, publication year and venue, and paper's abstract and research topics (see \sm{Detailed Dataset Description}).
We would like to note that there was a considerable data reduction in our data collection pipeline mainly due to missing content for the news stories and papers, but the data filtering affected different ethnic groups equally based on a chi-square test (\sm{Table S8}).

To quantify disparities in author mentions, we developed a computational method to identify three types of mentions in terms of author attributions, including name mentions, quotes, and institution mentions (see details in \sm{Check Author Attributions in Science News}).

We algorithmically inferred the perceived ethnicity and gender from authors' names, which mirrors how a reader might perceive social identities based on regularities in where the name originates. 
This choice may introduce bias because algorithmic inference may not be perfectly aligned with human perceptions \cite{kozlowski2021avoiding}, a limitation we return to in the discussion.

Overall, we do not measure the true ethnicity, but rely on the \textit{perceived} identity inferred from names. 
Therefore our conclusions should be interpreted as reflecting disparities among scientists with name-inferred ethnicities rather than self-identified ethnicities directly. 
However, we obtained consistent results when coding author names with racial self-identities using the U.S. census data or an alternate definition of ethnicity using the Wikipedia data (\sm{Fig. S4}).

We used mixed-effects logistic regression models to control for a broad range of plausible confounding factors, including corresponding author status, authorship position, author rank and popularity, name complexity, affiliation rank and location, abstract readability, team size, research topics, and news features such as year of coverage, article length, and journalist's demographics (see details in \sm{Mixed-Effects Regression Models}).
The model also adds random effects for publication venues and news outlets, enabling us to measure differential mentions within a particular news outlet covering a particular academic journal on a particular research topic.
This helps ensure that we are comparing media mentions of scholars doing comparable work. Nevertheless, our model cannot provide causal evidence of ethnic discrimination.

\section{Who Gets Mentioned Less Often in Media Coverage?}

We find substantial disparities in author mentions across name-inferred ethnicities---i.e., given that an author's research paper is being reported in the news, authors whose names are associated with certain ethnicities are much less likely to be mentioned by name. These disparities are robust to the inclusion of increasingly stringent controls of external factors that may influence the likelihood of being mentioned (Model 5 in \sm{Table S5}).
Specifically, compared to Anglo-named authors, most authors with minority-ethnicity\footnote{We refer to ``Anglo'' as the majority group based on the number of observations in our data (where all papers received media coverage). \sm{Table S2} shows that Anglo-named authors have their research covered more than twice as much as that of Western \& Northern European named authors, the second largest group.} names are significantly less likely to be mentioned, with European names disadvantaged the least while East Asian and African names disadvantaged the most.

In contrast to ethnicity, we find no disparity in author mentions across genders. However, when fixed effects for paper keywords are not considered, the author gender variable appears to have a significant effect (Model 3 in \sm{Table S5}). As gender representation varies widely across academic disciplines \cite{xie2003women,handelsman2005more}, this result suggests that gender differences in mention rates are likely to be explained by different mention rates across different fields. 

\begin{figure*}[t!] 
\centering
\includegraphics[trim=0mm 0mm 0mm 0mm, width=0.6\columnwidth]{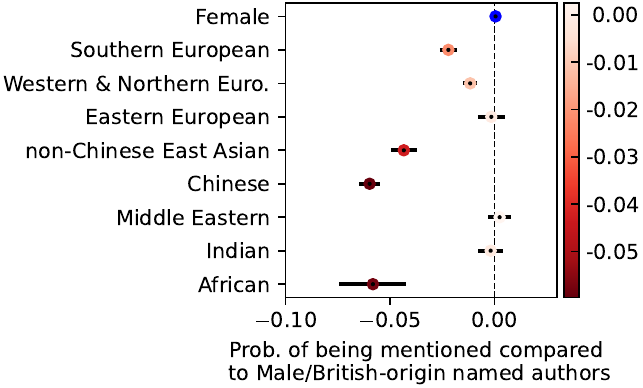}
\caption{The average marginal effects of author's gender and ethnicity on the probability of being mentioned by name in science news reporting their research. Estimations are based on our full model (Model 5) fitted to all 524,052 observations in our data.
Authors with names associated with minority ethnicities, especially East Asian and African names, are much less likely to be mentioned than Anglo-named authors after controlling for corresponding author status, affiliation rank and location, authorship position, author rank and popularity, last name complexity, abstract readability, team size, research topics, and news features such as year of coverage, article length, and journalist's demographics, as well as random effects for publication venues and news outlets.
Colors are proportional to absolute probability changes (legend is shown vertically for space consideration).
\textit{Woman} is colored as blue to reflect its difference from ethnicity identities.
The error bars indicate 95\% bootstrapped confidence intervals.}
\label{fig:main-bias-result}
\end{figure*} 

To quantify ethnic disparities in mentions, we calculated the average marginal effects for the author ethnicity and gender variables using the fullest model (Model 5 in \sm{Table S5}).
As shown in \fref{fig:main-bias-result}, the estimated probability of being mentioned is 1.2-6.0 percentage points lower for most ethnicities compared to the British origin.
As the average mention rate is only 41.2\% (see \sm{Detecting Author Name Mentions}), these absolute drops represent significant disparities: the 4.3-6.0 percentage points marginal decrease for East Asian and African names represents a 10.4\%-14.6\% relative decrease in media representation for authors with those names.
This result reveals that the U.S. mainstream media outlets exhibit profound disparity against non-Anglo-named authors in mentioning them by name in science news: Given the current disparities, \textit{we estimate that about six thousand minority-ethnicity scholars should have been mentioned in our data alone if they had Anglo names}.

\section{Large Disparities Still Exist for U.S.-based Authors with East Asian and African names}

\begin{table}[ht!]
\centering
\begin{tabular}{|l|d{5}|d{5}|r|}
\hline
\multicolumn{1}{|l|}{\textbf{Gender/Ethnicity}} & \multicolumn{1}{c|}{\textbf{U.S.-based}} & \multicolumn{1}{c|}{\textbf{non-U.S.}} & \multicolumn{1}{r|}{\textbf{p-value}} \\ \hline

Woman                       &	\cellcolor{red!1  } -0.01          & \cellcolor{green!1} 0.01 &	0.254 \\
Southern European            &	\cellcolor{red!1}   -0.02          & \cellcolor{red!33}	-0.33\sym{***} &	0.000 \\
Western \& Northern European &	\cellcolor{red!1}   -0.02          & \cellcolor{red!19}	-0.19\sym{***} &	0.000 \\
Eastern European             &	\cellcolor{green!12} 0.12\sym{***} & \cellcolor{red!24}	-0.24\sym{***} &	0.000 \\
non-Chinese East Asian       &	\cellcolor{red!23}  -0.24\sym{***} & \cellcolor{red!36}	-0.36\sym{***} &	0.003 \\
Chinese                      &	\cellcolor{red!31}  -0.31\sym{***} & \cellcolor{red!57}	-0.57\sym{***} &	0.000 \\
Middle Eastern               &	\cellcolor{green!7}  0.07\sym{**}  & \cellcolor{red!11}	-0.11\sym{***} &	0.000 \\
Indian                       &	\cellcolor{green!4}  0.04\sym{*}   & \cellcolor{red!16}	-0.16\sym{***} &	0.000 \\
African                      &	\cellcolor{red!29}  -0.29\sym{***} & \cellcolor{red!52}	-0.52\sym{***} &	0.034 \\ \hline
\end{tabular}
\caption{The regression coefficients of author's gender and ethnicity in predicting the author's name being mentioned in a news story reporting their research.
Disparities between minority and Anglo-named scholars are significant when they are all affiliated with international institutions, with each ethnicity reaching statistical significance. 
The disparities are reduced when scholars are all affiliated with U.S. institutions.
However, even among U.S.-based authors, there are significant disparities for authors with East Asian and African names, suggesting that location does not explain all disparities.
A separate Model 5 is trained for (i) the subset of U.S.-based institutions (ii) the subset of non-U.S. institutions. 
When fitting a model for the U.S. subset (or non-U.S. subset), we omitted the \textit{affiliation location} variable. 
Significance level: *** p$<$0.001, ** p$<$0.01, and * p$<$0.05. 
The p-values are based on the statistical test of differences in coefficients between two regression models using the equation provided in \cite{clogg1995statistical}.}
\label{tab:us-vs-non-us}
\end{table}

In science reporting, journalists often directly seek out the authors by phone or email to contextualize and explain their results. For authors at non-U.S. institutions, journalists from U.S.-based outlets could be less likely to reach out due to time-zone differences and communication costs associated with distance \cite{lin2022remote}, potentially resulting in a lower rate of being mentioned. 

Indeed, our regression model shows that international scholars are significantly less likely to be mentioned compared with their U.S. domestic counterparts of the same ethnicity (see the negative coefficient for affiliation location in \sm{Table S5}).
However, the significant negative coefficients for ethnicity in the same regression (\sm{Table S5}) suggest that location does not explain all disparities in mentions, as disparities between minority ethnicities and Anglo ethnicity still exist conditioning on authors being located in the same geographical region (whether inside or outside the U.S.).


We next quantified the disparities across different locations by measuring the size of disparities separately for (i) the subset of our data where the authors are all from U.S.-based institutions, (ii) that for all non-U.S. authors.

The coefficients in \tref{tab:us-vs-non-us} indicate that, \textit{among international authors, those with non-Anglo names are significantly less likely to be mentioned than those with Anglo names, despite that they are all distant to U.S.-based journalists}.
Compared to international researchers, the mention disparities are much smaller for U.S.-based authors, suggesting that being affiliated with a U.S. institution does decrease the disparity for each minority ethnicity, and for some groups including Indian, Middle Eastern, and Eastern European, the mention rate is even higher than Anglo-named authors.
Nevertheless, close proximity between journalists and authors does not eliminate all disparities in who is mentioned, \textit{as the disparities are still large and significant for African and East Asian ethnicities among all authors affiliated with U.S. institutions} (\tref{tab:us-vs-non-us}).

\section{Disparities Across Three Types of Mentions Among U.S.-based Authors}

\begin{figure*}[ht!] 
\centering
\includegraphics[trim=0mm 0mm 0mm 0mm, width=\linewidth]{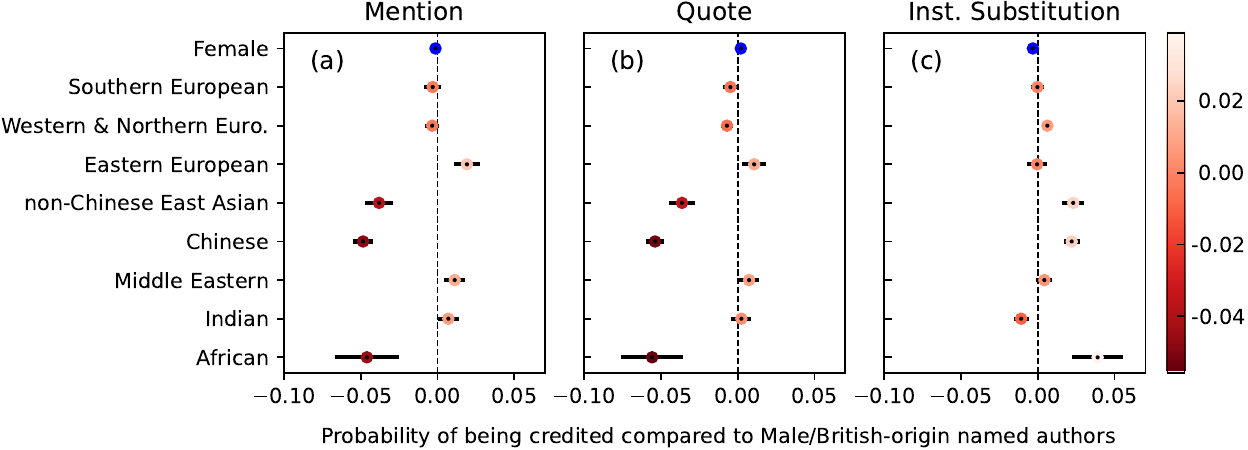}
\caption{U.S.-based authors with minority-ethnicity names are less likely to be mentioned by name (\textbf{a}) or quoted (\textbf{b}), and are more likely to be substituted by their institution (\textbf{c}). The average marginal effects are estimated based on our finest model (Model 5) fitted to 317,626 observations where the author is from U.S.-based institutions.
A negative (positive) marginal effect indicates a decrease (increase) in the probability of being credited compared to authors with Man (for gender) or Anglo (for ethnicity) names. The colors are proportional to the absolute probability changes.
\textit{Woman} is colored as blue to reflect its difference from ethnicity identities.
The error bars indicate 95\% bootstrapped confidence intervals.}
\label{fig:us-bias-result}
\end{figure*} 

Our prior result shows that ethnic disparities in mentions are observed even among authors based in the U.S., where scheduling difficulties and other pragmatic factors should be minimized or, at least, not associated with ethnicity. 
Focusing on U.S.-based authors, we further separated mentions into different types to better understand the mechanisms driving these disparities.
We first quantified the average marginal effects in mention rates based on a Model 5 fitted to the U.S. subset (\tref{tab:us-vs-non-us}).
\fref{fig:us-bias-result}a shows that U.S.-based Chinese, non-Chinese East Asian, and African-named authors experience a 4.8, 3.8, and 4.6 percentage points drop in mention rate compared to their Anglo-named counterparts. 

One plausible mechanism generating mention disparities for U.S.-based minority authors is journalists' perceptions of, or actual differences in, authors' fluency in speaking English.
While authors' actual fluency is not available to journalists, they may make assumptions about an author's speaking fluency based on ethnicity or other factors. 
For example, among U.S.-based authors, journalists may assume that authors with minority-ethnicity names are more likely to be foreign-born with less English fluency \cite{chiswick1998english,shields2002english}.
If so, journalists may be less willing to contact the author to ask them to explain the findings in cases where they need additional information to understand the paper, which could result in fewer quotations of these authors.

To measure disparities in quotation rates, we identified authors who are named as part of quotations (a subset of name mentions; see \sm{Author-Quote Detection}) and applied the same regression model to U.S.-based authors in predicting whether the author is quoted in the news story reporting their research. 
Fig~\ref{fig:us-bias-result}b shows that there are substantial disparities in quotation rates for authors with East Asian-associated and African-associated names.
\textit{We note that this result suggests, but does not prove, that the perceived fluency could be a driving mechanism, as other mechanisms such as the rhetorical value of names, may also produce this result}.

To more directly test the rhetorical mechanism, we examined ``institution-substitution'' where the author is mentioned by their institution but not by name (see \sm{Detecting Institution Mentions}), e.g., being named as ``researchers at the University of Michigan.''
Among U.S.-based authors, this mention type should not depend on pragmatic factors such as scheduling difficulties or perceived English fluency. Thus, this substitution effect likely reveals the rhetorical value journalists place on authors' names vs. institutions.

Fig.~\ref{fig:us-bias-result}c shows the probability of institution-substitution of minority-ethnicity authors relative to those with Anglo names, revealing that U.S.-based authors with African and East Asian names are more likely to have their names substituted for their institutions (\sm{Fig. S3} shows similar results of this analysis using the full data).
Analyzing three different types of author mentions thus reveal that while some mention disparities may be explained by perceived English fluency or other pragmatic factors, journalists' rhetorical choices are also key.


\section{Consistent Disparities Across Three Types of News Outlets}

\begin{figure*}[ht] 
\centering
\includegraphics[trim=0mm 0mm 0mm 0mm, width=\linewidth]{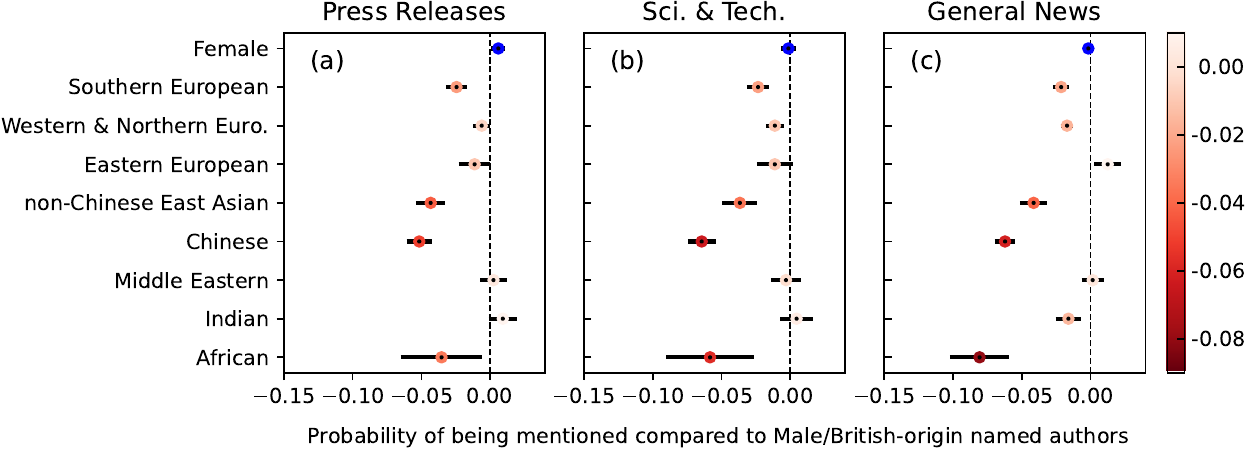}
\caption{Ethnic disparities in mention rates are consistent across three types of news outlets. The similar sizes of absolute disparities in three outlet types reflect starkly different magnitude of relative effects, as the average mention rates in Press Releases, Science \& Technology, and General News outlets are 63.5\%, 41.9\%, and 24.2\%, respectively.
In particular, the 6-8 percentage points drop in mention rates for Chinese and African named scholars reduces nearly one third of their expected media representation in General News outlets. 
We fitted a separate model (Model 5) to the subset of all observations in each type of outlet.
The colors are proportional to the absolute probability changes. Error bars represent 95\% bootstrapped confidence intervals.}
\label{fig:outlet-type}
\end{figure*} 

News outlets vary in the depth and breath of their reporting, e.g., Science \& Technology outlets write about 650 words per story on average, while General News outlets write about 900 words (\sm{Fig. S2}).
These differences suggest potentially important variability in the nature of journalists' day-to-day work and backgrounds.
To explore the discrepancy of disparities in author mentions across different types of outlets, we fitted the specification of Model 5 separately for three outlet types in our full data and quantified the average marginal effects.

\fref{fig:outlet-type} shows that ethnic disparities in mention rates surprisingly remain consistent across all outlet types, with authors of non-Anglo names being mentioned less likely. Larger disparities are found for ethnic categories that are more culturally distant from Anglo (e.g., East Asian and African).
Although the three outlet types have similar sizes of absolute disparities, they vary substantially in the relative scale, as the average mention rates of Science \& Technology outlets and General News outlets are 34.0\%-61.9\% less than Press Releases outlets (\sm{Table S4}). 

The disparity in Press Releases outlets is particularly notable, as stories in these outlets typically reuse content from university press-releases, suggesting that universities' press offices themselves, while less biased than other outlet types, still prefer to mention scholars with Anglo names.
This result is unexpected because local press offices are expected to have greater direct familiarity with their researchers, reduce the misuse of stereotypes, and to be more responsible for representing minority researchers equitably.

The largest disparities are seen in General News outlets, e.g., The New York Times and The Washington Post, where again scholars with Chinese- and African-associated names have 6.0-8.0 percentage points drop in mention rates. This significant drop reduces nearly one third of the deserved media representation of a large community of scientists, as General News outlets mention authors with a 24.2\% chance on average (\sm{Table S4}).
As General News outlets have well trained editorial staff and science journalists dedicated to accurately reporting science and tend to publish longer stories that have room to mention and engage with authors, this result is alarming.
Historically, these ethnic minorities have been stereotyped and underrepresented in U.S. media and leadership roles \cite{behm2013race, lu2020east}, which has continued in objective science reporting across all outlet types. 
The mechanisms of this variation deserve further investigation.

\section{Is the Situation Getting More Equitable?}

\begin{figure*}[ht!]
\centering
\includegraphics[width=0.5\columnwidth]{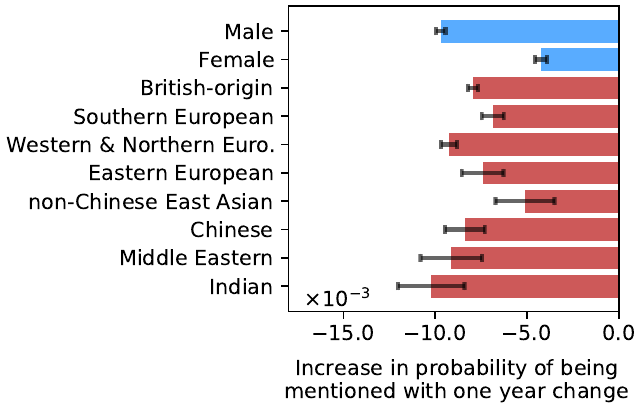}
\caption{Average marginal effects on the mention probability for a one-unit increase in the mention year variable for each gender (blue) and ethnicity (red). A separate model was fitted using all observations for each demographic group. The African ethnicity is not shown due to insufficient data for fitting the model. Error bars show 95\% bootstrapped confidence intervals.}
\label{fig:time}
\end{figure*}

The longitudinally-rich nature of our dataset allows us to examine how author mentions in science news have changed over the last decade. Mention rates are on average decreasing over time, as shown by the coefficient of the \textit{mention year} variable in Model 5 (\sm{Table S5}). 
To examine the time trends across demographic categories, a separate Model 5 was trained to quantify the marginal change in mention rate per year increase for each gender and ethnicity in our full data.
Note that demographic attributes not under study were still included in each model, e.g., when examining the temporal changes in mention rates for men and women, ethnicity was still included in the model, and vice versa. 

As shown in \fref{fig:time}, the mention year has a negative association with author mention rates for all gender and ethnic groups, and the larger decrease for Anglo indicates that their overall advantages are shrinking. 
Indeed, authors with non-Chinese East Asian names, one of the most disadvantaged group in this study, have the lowest decreasing rate.

However, the estimated rates of change are relatively small for most ethnic groups, suggesting that the existing disparities are unlikely to disappear in the short term without intentional behavior change. 
We refrain from making predictions as to when the mention equality will be reached eventually, or adopting sophisticated time series models to forecast the trajectory of mention rates in the long run, because such extrapolation will be of little practical use, especially given that the long-term changes in the academia and media practices remain unforeseeable. 


\section{Discussion}
\label{sec:discussion}

Our analyses reveal that the attention researchers get in science news reporting is strongly related to the ethnicities associated with their names. The effects are robust to a variety of plausible confounds, and even appear when controlling for the (1) particular news outlet, (2) particular scientific venue, and (3) particular research topic. Although we cannot claim that the reported effects are causal, this unusually strong observational evidence deserves further attention.

\subsection{Ethnicity and Gender}

Authors with most non-Anglo names are mentioned substantially less when their research is covered in U.S. science news. Mention rates are especially low for East Asian and African names, less pronounced for European names, are even less pronounced for Indian and Middle Eastern names.
The large disparity for East Asian ethnicities and its sharp contrast with South Asian categories (such as Indian) is consistent with the ``bamboo ceiling'' phenomenon observed in leadership roles in the U.S. \cite{lu2020east}. 
As science becomes more global and is increasingly driven by authors of non-Western ethnicities, the way English-language media responds to non-British-named scholars will only grow in importance. 
In contrast to ethnicity, we do not find gender disparities in mentions of scholars once the research fields are controlled for. One possible reason is that fields vary in their overall level of mention rates and in their gender representation \cite{handelsman2005more}. Looking within fields masks gender disparity that may exist between them.

\subsection{Ruling in and out different mechanisms}

Our analyses point to a multi-causal generation of ethnic disparities, in which both pragmatic difficulties of interviewing researchers (location and possibly perceived fluency) and journalists' tastes regarding names' rhetorical values play key roles. 

In support of pragmatic difficulties, we find that international location (which hosts more scholars with non-Anglo names) has a negative effect on mention rates.
However, location is not the driving mechanism as 
disparities persist among both international authors and U.S.-based authors, which would disappear if location was the decisive factor.
In support of English fluency, we find that ethnic disparities for East Asian and African ethnicities appear in quotations among U.S.-based authors, who are unlikely to suffer from pragmatic difficulties in scheduling interviews, but may differ in their actual or perceived fluency. However, we note that these disparities may be produced by other factors such as assertiveness \cite{lu2022surprising}.

In addition to these pragmatic factors, journalists' rhetorical choices are key.
In support of this mechanism, journalists are more likely to ``substitute'' a direct name mention with the researcher's institution for authors with East Asian and African names, suggesting that the context of discovery is important, but the institution serves the journalists' rhetorical goals better than the name.
Additional evidence comes from outlet types: when journalists' role in the news articles is minimal---when the outlet simply republishes a university press release---the relative disparities are also minimal; when the news stories are written by journalists themselves, the relative disparities are the largest. 
However, we note that the disparities in Press Releases outlets also suggest that journalists are not the only actor behind the inequality.

The data do not allow us to fully explain journalists' rhetorical choices. For example, we hypothesize that their choices may be driven by journalists' own perceptions of author's authoritativeness or by expected tastes of their audiences. However, we observe that ethnic disparities in mentions do not vary substantially across Science \& Technology and General News outlets, although the two likely differ in their audiences. This observation suggests again that journalists' personal preferences play an important role. Furthermore, in \sm{Section IV.B}, we examine whether the interaction between authors' and journalists' name-inferred identities are associated with the mention rate, but do not find clear evidence. Disentangling the source of journalists' choices is an important avenue for future work.

\subsection{Limitations}

Although the scale and the breath of our dataset enable the use of unusually fine-grained controls, the analysis is not without limitations. First, the observational nature of the data precludes strong causal statements.
Second, the analysis was conducted with perceived ethnicities, which do not reflect self-identities accurately, nor account for multi-ethnicity identities.
We hope our work stimulates the collection of such data where possible, to enable more accurate and fine-grained conclusions \cite{nyt2020scientific}.
Thus, a key limitation of our design and the voluminous audit study literature must be acknowledged: such types of studies do not measure whether journalists actually form an inference of ethnicity when seeing names. We believe assuming that they do form such inferences is very reasonable and supported by the large empirical disparities we observe here. 
More direct evidence on journalists' decision processes is a fruitful direction for future research.
Besides, we inferred the perceived ethnicity via a name-based classifier, \textit{Ethnea}. Although journalists, like the classifier, may have no information about authors except their names, the inference will undoubtedly not match all actual human perceptions about the authors.  
Furthermore, the classifier is unable to identify key demographic groups, such as African American scholars. 
Nevertheless, as an exploratory test, we repeated our analysis using a classification of race based on the U.S. census data (\sm{Fig. S4}), which includes ``Black'' as one of the labels. The result does not show statistically significant under-representation of Black scholars relative to ``White.''
Note that African-named authors (based on \textit{Ethnea}) are not necessarily classified as ``Black'' based on the Census data (\sm{Tables S6-S7}). 

Third, some plausible covariates are unavailable for inclusion, such as the number of citations a paper received at the time of being mentioned. However, we anticipate the effect of such covariates to be small given current controls. Furthermore, the majority of papers in our data were mentioned within one year after publication (\sm{Fig. S1}), which limits the number of citations a paper can accrue in such a short time window.
Relatedly, we lacked direct measure of authors' English fluency, and our measure of prestige-related factors such as author rank and affiliation rank may not be able to accurately reflect prestige.

Fourth, we did not test other potential mechanisms. For instance, reporters often choose to interview authors listed as the corresponding author and prestigious authors from top institutions. Although our model controls for the corresponding author status, author rank, and affiliation rank, we did not examine how much of the disparities are driven by these prestige-related factors.
It is possible that which author of a paper is designated as corresponding and their selection into top institutions are themselves a product of structural discrimination with respect to authors' demographics.
Thus disparities observed in press mentions may be partly driven by decades or centuries of decisions that are ingrained in institution hiring practices \cite{small2020sociological}.

Fifth, our data contain too few examples of some ethnicities (e.g., Polynesian and Caribbean) to accurately estimate disparities; such ethnicities are regrettably omitted, though we recognize that these groups likely experience disparity from their minority status as well.

Sixth, our study has focused solely on U.S.-based news outlets. Many of these outlets are often global in reach and mentions in them often serve as markers of prestige for scholars. However, behaviors of these outlets may not be representative of broader media reporting practices. At present, the Almetrics data only provide sufficient quantity for U.S.-based outlets that allows us to control for potential confounds related to each outlet (62\% of all news mentions are solely from U.S.-based outlets), which is critical for our study design. Nevertheless, bias is likely not unique to one country and additional global scale data is necessary to move beyond a U.S. focus and study country-specific and global journalistic practices.

Lastly, this research relies on large-scale datasets and algorithms that may themselves encode systemic social inequalities. For instance, which venues are considered ``mainstream'' and therefore worthy of tracking by Altmetric may be the outcome of racial inequities \cite{alamo2020racialization}. Which groups the algorithms choose to identify as distinct groups are choices that may reflect long histories of racialization seen through a ``white racial frame'' \cite{tatum2017all,feagin2020white}. 
The availability of data also drove our focus on English-language science and media, thereby accumulating more activity around certain cultures than others. We believe these limitations place substantial scope conditions on the findings.

\subsection{Conclusions and Implications}

Our work shows that science journalism is rife with disparities in which author receives name attribution, with authors from certain ethnic group receiving much more name mentions and quotations than their peers when their comparable research papers are all reported in U.S. news.
These ethnic disparities likely have direct negative consequences for the careers of unmentioned scientists, and skew the public perception of who a scientist is---a key factor in recruiting and training new scientists. 

Our findings have two implications for science policy and science journalism.
First, bringing the attention to large-scale ethnic disparities in author mentions in science news, of which journalists may themselves have been unaware, can be an agent of change. 
Second, decision-makers at U.S. research institutions may take these ethnic disparities into account when making hiring or promotion decisions. 
More importantly, addressing this problem requires more research to investigate the mechanisms leading to it, which we hope this paper helps stimulate.

While our study only focuses on the ``second stage'' of science media coverage -- its quality -- it is likely that such ethnic disparities would be even larger in the first stage of coverage where media outlets choose whose papers to report on in the first place. 
Supporting evidence comes from recent empirical studies that find gender and regional disparities in the online attention to scientists' work \cite{vasarhelyi2021gender, chapman2022altmetric, davidson2023analysis}. 
Our work thus suggests that disparities in science media are likely to compound across different aspects of coverage, yielding ultimate differences in outcomes much larger than those shown by studies of any one stage.


\bibliographystyle{unsrtnat}
\bibliography{references}

\vspace{10mm} 

\noindent \textbf{Acknowledgements:} We thank \textit{Altmetric.com} for sharing the mention data used in this study. This work uses Web of Science data by Clarivate Analytics provided by the Indiana University Network Science Institute and the Cyberinfrastructure for Network Science Center at Indiana University. We thank Aparna Ananthasubramaniam, Jiaxin Pei, Daniel M. Romero for helpful discussions and suggestions. \textbf{Funding:} This work has received no funding. \textbf{Author contributions:} H.P., M.T. and D.J. collaboratively conceived and designed the study. H.P. performed the analyses. H.P., M.T. and D.J. drafted and revised the final manuscript. \textbf{Competing interests:} The authors declare that they have no competing interests. \textbf{Data availability:} All code used in this study is available at: \url{https://github.com/haoopeng/author_mentions}. The Altmetric data of news coverage is available free of charge to university-affiliated scientometric researchers; see details at: \url{https://www.altmetric.com/research-access/}. The Microsoft Academic Graph data can be publicly accessed at: \url{https://openalex.org/}.

\subsection*{Supplementary Materials}

SI Data and Methods\\
SI Tables S1 to S11\\
SI Figures S1 to S4\\
Supplementary Text

\end{document}


\title{Author Mentions in Science News Reveal Widespread Disparities Across Name-inferred Ethnicities \\ (Supplementary Information)}
\author{Hao Peng, Misha Teplitskiy, David Jurgens}


\date{\today}

\maketitle

\section{Data and Methods}

\subsection{Detailed Dataset Description}

\subsubsection{News Stories Mentioning Research Papers}

The dataset of news stories mentioning scientific papers was collected from \textit{Altmetric.com} (accessed on Oct 8, 2019), which tracks a variety of sources for mentions of research papers, including coverage from over 2,000 news outlets around the world. 
To control for differences in the frequency of scientific reporting and potential confounds from variations in journalistic practices across different countries, the list of news outlets was curated to 423 U.S.-based news media outlets, with each having at least 1,000 mentions in the Altmetric database. Location data for each outlet is provided by Altmetric.
This exclusion criterion ensures that the dataset has sufficient volume to estimate outlet-level disparities, while still retaining sufficient diversity in outlet types, stories, and the scientific articles they cover.
%
Each mention in the Altmetric data has associated metadata that allows us to retrieve the original citing news story as well as the DOI for the paper itself.

\subsubsection{Scraping News Content and Identifying Journalists} \label{sec:journalist}

Due to access and permission limitations when retrieving the content of news stories, 135 outlets were excluded due to insufficient volume (27 outlets denied our access entirely; 65 outlets had less than 100 urls crawled; 43 outlets had at least 100 urls crawled, but only with non-news content such as subscription ads). For the remaining 288 outlets, 44.1\% of the stories were successfully retrieved after cleaning, including dropping duplicated htmls and removing all html tags and unrelated content such as advertisements. Stories with less than 100 words were removed (less than 1\%) as a manual inspection showed that the vast majority of these do not contain the complete content of the story. This process resulted in 520,061 downloaded news stories mentioning 275,403 papers from the 288 outlets.

In order to control for the effects of journalists' ethnicity and gender, we first used the \textit{newspaper} Python package (\url{https://github.com/codelucas/newspaper}) to extract the journalists' names from the retrieved html news content. 
Since not all stories in each outlet contain the journalist information and the \textit{newspaper} package does not work perfectly for every story that has journalist information, we focused on the top 100 outlets (ranked by the story count). 
With manual inspection, we verified that this package can consistently and reliably identify journalists' names for 41 of the top 100 outlets.
We excluded extracted names with words signaling institutions and organizations (such as ``University'', ``Hospital'', ``World'', ``Arxiv'', ``Team'', ``Staff'', and ``Editors''). 
We also cleaned names by removing prefix words, such as ``PhD.'', ``M.D.'', and ``Dr.''.
We eventually obtained the journalist's name in 100,163 news stories (18.1\% of all cleaned stories) for 41 outlets. Note that we did not drop any data where the journalist's name is missing. When coding journalists' gender and ethnicity, we assigned ``Unknown'' to those missing names.

\subsubsection{Retrieving Paper Metadata} \label{sec:mag}

The Altmetric database does not contain detailed author information and therefore an additional dataset is needed to identify the authors of mentioned papers. 
We used the Microsoft Academic Graph (MAG) data \cite{sinha2015overview} (accessed on June 01, 2019) to retrieve information for each paper based on its lower-cased Document Object Identifier (DOI). 
%
%
MAG also provides rich metadata for papers, including author names, author rank, author affiliations, affiliation rank, publication year, publication venue, the paper abstract, and paper topical keywords. 
Although most papers in the Altmetric database are also indexed in MAG, it does not provide complete coverage for some important paper-level features such as the paper's abstract. 
As our regression models consider many control variables, we excluded papers with missing information on these variables, leaving us with 100,486 papers reported by 223,587 news stories in the final dataset.

\subsubsection{Obtaining Corresponding Authors} \label{sec:triplet}

We further used the Web of Science database (2019 version) to retrieve the corresponding author information for 86.0\% papers in the final dataset based on the DOI. 
We coded the corresponding author status as ``Unknown'' for the remaining missing papers, which are mainly from disciplines such as computer science that do not have the norm to designate corresponding authors.

\subsubsection{Final Dataset of Story-Paper-Author Triplets} \label{sec:final-data}

To examine whether a specific author is mentioned, we treated each (\textit{story, paper, author}) triplet as an observation in the regression.
We focused on several authors whom journalists are likely to mention by name when covering a paper in a news story, \textit{including the first author, the last author, and any middle author who is designated as the corresponding author}(note that the first author and the last author can be corresponding as well). 
It is possible that some papers could have equal-contributing first authors, however, our data does not have this information. 
We estimate that such cases are rare.
For solo-author papers, we included the single author in the analyses. 
Papers in a few research fields that commonly use the alphabetic-based authorship ordering are also included as journalists may be unfamiliar with this norm.

The final dataset consists of 223,587 news stories referencing 100,486 research papers. The data reduction from our initial source is mainly due to missing full text for news stories and missing important metadata for papers, especially the abstract. However, a chi-square test shows that the ethnicity distribution after data reduction (shown in Table~\ref{tab:before_after}) is almost identical to that before filtering (p = 1.0). 

As some stories mentioned more than one paper and some papers were mentioned in more than one story, we have 276,202 (story, paper) mention pairs. Since multiple authors are likely to be mentioned per paper, we have 524,052 (story, paper, author) triplets in total to test whether an author is mentioned in a story.

The distribution of the number of papers and news stories over time and attention per paper are shown in Figs.~\ref{fig:count}a-b. News story data is left censored and primarily includes stories written after 2010, as \textit{Altmetric.com} was only launched in 2012, which limits the collection of earlier news.
As shown in Fig.~\ref{fig:count}c, news stories can mention papers that were published several decades before, highlighting the potential lasting value of scientific work. However, the majority of papers are mentioned within the same year or just a few years after publication.
Table~\ref{tab:ethnea-broad-paper-mention} shows the the number of authorships and triplets for authors in each broad ethnicity group, and Table~\ref{tab:ethnea-broad-journalist} shows the number of triplets by journalists' inferred ethnicities.

\subsection{News Outlets Categorization} \label{sec:outlet-cate}

To estimate differences across outlets, we grouped 288 news outlets into three categories according to their news report publishing mechanisms (Table~\ref{tab:outlet-list}). The three categories are (1) Press Releases, (2) Science \& Technology, (3) General News.
The categorization is based on manual inspections of three random stories per outlet.

\begin{figure}[h!] 
\centering
\includegraphics[trim=0mm 0mm 0mm 0mm, width=0.35\columnwidth]{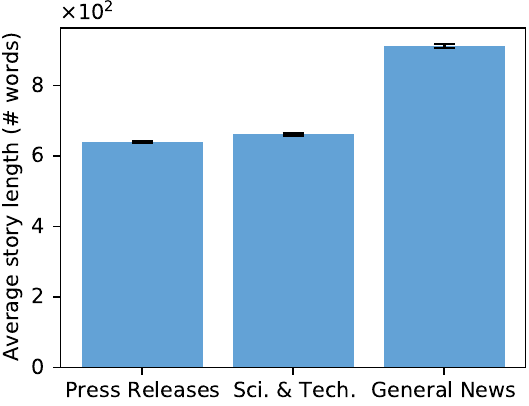}
\caption{The average story length for three types of outlets. Error bars show 95\% confidence intervals.}
\label{fig:story-length}
\end{figure}

The Press Releases category is unique since many outlets in this group commonly---if not exclusively---republish university press-releases as stories, making them reasonable proxies for estimating disparity in universities' own press office.
The Science \& Technology category consists of magazines that focus on reporting science, such as ``MIT Technology Review'' and ``Scientific American.'' These outlets typically construct a large scientific narrative referencing several papers in their stories.
The General News category includes mainstream news media such as ``The New York Times'' and ``CNN.com'' that publish stories in a wide variety of topics. They have well-trained editorial staff and science journalists who are focused on accurately reporting science.
Table~\ref{tab:outlet-type-avg-rate} shows the number of (story, paper, author) triplets by outlet types. The average number of words per story for each outlet type is shown in Fig.~\ref{fig:story-length}.

\subsection{Inferring Author and Journalist Gender and Ethnicity} \label{sec:ethnea}

As authors' gender and ethnicity are not directly available, we relied on the inferred demographic associations of their name. While such inferences could be inaccurate relative to how authors self-identify, self-identities are generally not available to journalists. 
Instead, classifier-based predictions on gender and ethnicity reflect stereotypical norms of the \textit{expected} demographics given a name---norms that journalists are likely to share and unconsciously use when first examining the author names of a paper and deciding whom to mention. Therefore, while imperfect, we based our study on these inferred attributes.

Gender and ethnicity were inferred using the \textit{Ethnea} API \cite{torvik2016ethnea}, which is specifically designed for use in bibliometric settings like ours. 
The library makes its prediction based on the nearest-neighbor matches on authors' first and last names using the PubMed database of scholars' country of origin, which offers superior performance over alternative approaches \cite{ambekar2009name,treeratpituk2012name}.

Author names in the MAG have varying amounts of completeness. While most have the first name and surname, special care was taken for three cases:
(1) If the name has a single word (e.g., Curie), the ethnicity and the gender were both set to \emph{Unknown}, as \textit{Ethnea} requires at least an initial. Single-word name cases occurred for 208 authorships in the final dataset.
(2) If the name has an initial and surname (e.g., M. Curie), we directly fed it into the API, which provides an ethnicity inference but returns \emph{Unknown} for gender due to the inherent ambiguity.
%
(3) If the name has three or more words, we took the first word as the given name and the last word as the surname. However, if the first word is an initial and the second word is not an initial, we took the second word as the given name (e.g., M. Salomea Curie would be Salomea Curie) to improve prediction accuracy and retrieve a gender inference.

While \textit{Ethnea} is trained with scholar names, we also applied it to infer the gender and ethnicity of journalists.
\textit{Ethnea} assigns fine-grained ethnic categories that are leaning towards country of origin.
Here, we recognize that ethnicity, race, and nationality are three related concepts.
Ethnicity categorizes people based on origin and cultural background, which is often reflected in names, whereas race is a social construct. 
In contrast, nationality reflects country of affiliation and is fluid due to immigration or migration.
We thus used the term ``ethnicity'' because it is the most accurate and relevant concept in the study of names.

To test for macro-level trends around larger ethnic categories and to ensure sufficient samples to estimate the effects, we grouped the 24 observed individual ethnicites from \textit{Ethnea} into 9 higher-level categories based on geographical proximity and cultural distances (Table~\ref{tab:ethnea-broad}), including (1) African, (2) Anglo, (3) Chinese, (4) non-Chinese East Asian, (5) Eastern European, (6) Indian, (7) Middle Eastern, (8) Southern European, (9) Western \& Northern European.

Note that due to the sample size and our hypotheses, \textit{African}, \textit{Chinese}, \textit{Indian}, and \textit{English} (renamed as ``Anglo'') were kept as separate high-level categories. \textit{Caribbean} and \textit{Polynesian} authors were excluded due to less than 100 observations (triplets) in total.
A few authors with organization names were also excluded.
Examples of names classified into each ethnicity are provided in Table~\ref{tab:eth-names}.
\textit{Ethnea} returns binary gender categories: \textit{Female} and \textit{Male}, though we recognize that researchers may identify with gender identities outside of these two categories. 
For both gender and ethnicity separately, some names are classified as ``Unknown'' if no discernible signal is found for the respective attribute by \textit{Ethnea}.

\subsection{Check Author Attributions in Science News}
\label{sec:check}

We developed a computational approach to identifying author mentions, quotes, and institution mentions for each (story, paper, author) triplet.

\subsubsection{Detecting Author Name Mentions}
We normalized both the news content and the author names to ensure that this approach works for names with diacritics. 
For each story-paper-author triplet, the author's last name was searched for using a regular expression with word boundaries around the name, requiring that the name's initial letter be capitalized.
While the chance exists that this process may introduce false positives for authors with common words as last names (e.g., ``White''), such cases are rare because (i) few authors in our dataset have common English words as their last names, and (ii) these words rarely appear at the beginning of a sentence in the story when they would be capitalized. 
However, a particular exception is for two common Chinese last names ``He'' and ``She,'' which can appear as third person pronouns at the start of sentences. 
We thus imposed additional constraints for these two names such that they must be immediately preceded with one of the following titles to be considered as a name mention: ``Professor'', ``Prof.'', ``Doctor'', ``Dr.'', ``Mr.'', ``Miss'', ``Ms.'', `Mrs.''.
Occasionally, the author name can occur within a reference to the paper at the end of the story, which should not be counted as a name mention. 
As authors are typically mentioned at the beginning or in the middle of the news story, we removed the last 10\% of the story content when checking name mentions (note that we obtained similar results without this filtering). 
Ultimately, author names were found in 41.2\% of all (story, paper, author) triplets.

\subsubsection{Author-Quote Detection}

Authors can be mentioned by name in different forms, including quotation (e.g., ``We are getting close to the truth.'' said Dr. Xu.), paraphrasing (e.g., Timnit says she is confident, however, that the process will soon be perfected.), and simple passing (e.g., A recent research conducted by Dr. Jha found that drinking coffee has no harmful effects on mental health.). 

We used a rule based matching method to detect explicit quotes for each (story, paper, author) triplet. We first parsed our news corpus using \textit{spacy} (\url{https://spacy.io/}). We identified 18 verbs that were commonly used to integrate quoted materials in news stories, from the most 50 frequently used verbs in our news corpus, including ``describe'', ``explain'', ``say'', ``tell'', ``note'', ``add'', ``acknowledge'', ``offer'', ``point'', ``caution'', ``advise'', ``emphasize'', ``see'', ``suggest'', ``comment'', ``continue'', ``confirm'', ``accord''. 
A sentence is determined to contain a quote from the author if the following two conditions are met: (i) both the quotation mark and the author's last name appear in the sentence, and (ii) any of the 18 quote-signaling verbs (or their verb tenses) appears within five tokens before or after the author's last name. A manual inspection of 100 extracted quotes revealed no false quote attributions.
This method only gives an underestimation of the quote rate, as it may not be able to detect every quote due to unusual writing styles or article formatting. So the advantage of Anglo-named scholars in being quoted may be even higher. 

\subsubsection{Detecting Institution Mentions}

We checked institution mentions based on exact string matching with authors' listed institution names in the MAG, i.e., for each (story, paper, author) triplet, we examined whether any of the author's full institution name appears in the news story. Similar to quote detection, this method may not be able to identify every instance of institution mentions due to noise in the MAG or the story using slightly different nomenclature such as an institution's abbreviation. However, a full list of alternative names for each institution is not available to us, we thus used this conservative method. For this reason, minority scholars' trend in being substituted by institutions is likely  underestimated. 

\subsection{Mixed-Effects Regression Models}


We adopted a mixed-effects logistic regression framework to examine the demographic disparity in author mentions in science reporting. 
In our regression framework, each (story, paper, author) triplet is an observation, with the dependent variable indicating whether the author is mentioned or not in the story. 
Many factors are known to influence name mentions that could confound the analysis of ethnicity and gender, such as author reputation, institutional prestige and location, publication topics and venues, outlets, and journalist demographics.
Here, we provide details of these factors and present a series of five regression models that build upon one another by adding more rigorous control variables at each step.
The increasing level of model complexity allows us to test the robustness of the effects of ethnicity and gender association, and also to examine potential factors at play in science coverage.

\subsubsection*{Model 1: Naive Disparity}

The first model directly encodes our two variables of focus, gender and ethnicity, as the sole categorical factors in the regression. Here and throughout the study, we treat the reference group for ethnicity as \textit{Anglo} and for gender as \textit{Man}. While overly simplistic in its modeling assumptions, Model 1 nevertheless tests for systematic differences for whether authors of a particular demographic are mentioned less frequently and serves as a baseline for layering on controls to explain such disparity.

\subsubsection*{Model 2: Paper Author Controls}

Many author-level attributes other than demographics could influence journalists' perceptions on authors and the coverage of them. Model 2 introduces 13 additional factors to control for features of papers' authors. 

\textit{Prestige Factors.}
The reputation of the author may also influence the chance of being named. High-status actors and institutions tend to receive preferential treatment within science \cite{merton1968matthew,azoulay2013matthew,tomkins2017reviewer}, and we hypothesize that these prestige-based disparities may carry over to media coverage as well. 
To account for prestige effects, we include the author rank and institution rank provided by the MAG \cite{wang2019review}.
We take the highest institution rank for authors with multiple affiliations.
This ranking estimates the relative importance of authors and institutions using using a heterogeneous citation network derived from metadata of all published papers in the literature.
This weighted metric has been shown to produce more fine-grained and robust measurements of impact and prestige, and it's less open to manipulation than traditional coarse measures such as h-index and citations \cite{wang2019review}.
Institution and author ranks are not necessarily directly related, as institutions may be home to authors of varying ranks (e.g., early- or late-career faculty) and the same author may appear with different affiliations on separate papers due to a career move. 
Note that for \textit{rank} values, negative-valued coefficients in the regression models would indicate that higher-ranked individuals and those from higher-ranked institutions are more likely to be mentioned.

We also add a variable indicating the author's institution location with three categories: (1) domestic, (2) international, (3) unknown. For authors with multiple affiliations, we assign ``domestic'' if there is at least one U.S. institution.
This variable controls for geographical factors that may influence journalists' willingness to contact by phone or video chat service and therefore influence whether they mention the author. We infer the country for institutions based on their latitude and longitude provided in the MAG.

Popular authors who have lots of press coverage may be more likely to be mentioned. We add a factor indicating whether the author is among the top 100 most popular scholars based on their number of papers mentioned in the news in our final dataset. 

In multi-author papers, the team often designates one or more corresponding authors, who are presumably more likely to be contacted and therefore mentioned by journalists. 
Our data includes the corresponding author information for most papers.
We thus include a variable indicating whether the author is corresponding or not on the covered paper.

\textit{Last Name Factors.}
People are known to have a preference for both familiar and more easily-pronounceable names \cite{song2009if,laham2012name}, and this preference could potentially affect which author a journalist mentions. Therefore, we introduce two factors as proxies: (1) the number of characters in the last name as a proxy for ease of pronunciation, and (2) the log-normalized count of the last name per 100K Americans from the 2018 census data. As journalists are drawn from U.S.-based news sources, the latter reflects potential familiarity. 

\textit{Other Authors.}
Scientific knowledge is increasingly discovered by teams, as tackling complex problems often require the collaboration between experts with diverse sets of specialization \cite{guimera2005team,greene2007demise,milojevic2014principles}.
On these multi-author projects, the first authors are commonly junior scholars who are directly responsible for the work; 
the last authors are typically the senior author responsible for directing the project; these two author positions are suggested in science journalism guidelines when determining whom to interview \cite{blum2006field}.
We thus control for the position of an author with four categories: (1) first position, (2) middle position, (3) last position, (4) solo author. The last author position is used as the reference category in the regression. 

When journalists examine a paper's authors, the team size may influence their understanding of the distribution of credits among authors, potentially reducing the chance of any author being mentioned for papers with many authors.
We thus include a variable for the number of authors.

\subsubsection*{Model 3: Paper and Story Content}

The content of the paper and story, and journalist demographics also can play a role in affecting author mentions. We thus control for the following factors in Model 3.

\textit{Year of News Story (Mention Year).}
Disparities in science coverage may have temporal variations due to unpredictable factors that are directly or indirectly related to research.
For instance, the available funding resources can affect the number of research outputs in a year, which would in turn influence the amount of time and space journalists devote to scientists in news articles.
We thus control for the year of the news story, i.e., the mention year of the paper. We treat it as a scalar variable (zero-centered). 

\textit{Year Gap between Story and Paper.}
News stories often reference older scientific papers in the narrative, as shown in Fig.~\ref{fig:count}c. For older papers, at the time of a recent story publication, the original authors may be unable to be reached or the story may be framed differently from recent science that is considered ``fresh.'' Indeed, citing timely scientific evidence in a news report can increase  credibility perceptions of a story \cite{sundar1998effect,rieh1998understanding}. Therefore we include a variable that quantifies the year difference between the mention year and the publication year of the mentioned paper.

\textit{Number of papers mentioned in a story.}
A story can mention several papers to help frame and construct its scientific narrative, and potentially increase its news credibility perception. However, referencing many papers in a story may reduce the amount of space and attention allocated to each paper by journalists, and therefore may decrease the chance of its authors being mentioned. We thus control for the number of mentioned papers in a story. 

\textit{News Story Length.}
Longer articles provide more space in depicting stories about the science being covered, we thus control for the story length, measured as the total number of words.

\textit{Paper Readability.}
Given the tight timelines under which journalists work, quickly identifying and understanding insights is likely critical to what is said about a paper. A paper's readability may thus influence whether a journalist feels the need to reach out to the author, with more readable papers requiring less contact.
Readability, in turn, may also be tied to author's demographics like gender \cite{hengel2017publishing}, making it important to take readability into account.
%
Due to licensing restrictions, the full text of the majority of papers is unavailable freely; therefore we compute readability over the paper abstract using three factors: (1) the Flesch-Kincaid readability score, which estimates the grade-level needed to understand the passage; (2) the number of sentences per paragraph, which is a proxy for information content and density; and (3) the type-token ratio, which is a measure of lexical variety.
Another reason we focus particularly on the abstract is that journalists may not read the entire paper but very likely read the abstract.

\textit{Journalist Demographics.}
It is ultimately the journalist's decision to mention authors when writing science reports. Motivated by the commonly observed homophily principle in social networks \cite{mcpherson2001birds}, we hypothesize that the mentioning behavior in science reporting is associated with homophilous effects by ethnicity and gender.
To model such effects, we include the journalists' demographics in the regressions.
Due to insufficient instances of journalists identified in news stories (Table~\ref{tab:ethnea-broad-journalist}), we further coarsen the 9 broad ethnicity categories into four groups: (1) Asian (Chinese, Indian, and non-Chinese East Asian), (2) Anglo, (3) European (Eastern European, Southern European, Western \& Northern European), and (4) Other/Unknown (Middle Eastern, African, and Unknown).

\subsubsection*{Model 4: Paper Domains and Topics}

Some scientific domains and topics may be inherently more attention-getting than others. Some may be harder to understand without seeking additional explanation from authors. Furthermore, journalists' academic backgrounds may be unequally distributed across scientific fields, resulting in different propensities to reach out to authors. 

We thus include factors to capture the paper's topics using data from the MAG, which includes a large volume of keywords (665K) at different levels of specificity. A paper can have multiple keywords, with each having a confidence score between 0 and 1.
To capture high-level topical and methodological differences, we focus on the most common 199 keywords that occur in at least 500 papers in our final dataset.
Each keyword is used as an independent variable in the regression, whose value is the keyword's confidence score for the paper. 
 
\subsubsection*{Model 5: News Outlets and Publication Venues} 

Individual news outlets may follow different standards of practice in how they describe science, creating a separate source of variability in who is mentioned.
Publication venues each come with different levels of impact and topical focus that potentially affect the depth of journalistic focus on papers published in them.
%
To accurately model these sources of variations, we treat outlets and venues as \textit{random effects} in regression Model 5.
This mixed-effect regression model implicitly captures a robust set of factors involved in science reporting such as the tendency of specific journals to be mentioned more frequently (e.g., \textit{Nature}, \textit{Science}, or \textit{JAMA}) and the focus of news outlets on specific topics covered by different journals.

\subsection{Additional Ethnicity Coding}
\label{sec:validation}


Although \textit{Ethnea} is specifically designed for inferring scholars' ethnicity in bibliographic records, it is not expected to be entirely error-free.
As a robustness check, we replicated our analyses by inferring the ethnicity for the names of authors and journalists using two separate data sources to test whether the observed disparity persists. 

Specifically, we used the \textit{ethnicolr} (\url{https://pypi.org/project/ethnicolr/}) library to code ethnicity using either data derived from (i) the nationalities listed in Wikipedia infoboxes to infer nationality-based ethnicity, or (ii) self-reported ethnicity data associated with last names from the 2010 U.S. census. 
While these two sources of data use different definitions and granularities of ethnicity from \textit{Ethnea}, they nonetheless provide approximately-similar categories to \textit{Ethnea} that enable us to validate our results.

\subsubsection{Ethnicity based on Wikipedia}

We used the Wikipedia infobox data to code ethnicity based on the first name and the last name  \cite{ambekar2009name, sood2018predicting}. 
To make the results comparable to that based on \textit{Ethnea}, we placed 13 individual ethnicities defined in the Wikipedia into 8 broad categories: 

\begin{itemize}
    \setlength\itemsep{-0.5em}
    \item (1) African (\textit{Africans}),
    \item (2) Anglo (\textit{British}),
    \item (3) East Asian (\textit{EastAsian}, \textit{Japanese}), 
    \item (4) Eastern European (\textit{EastEuropean}),
    \item (5) Indian (\textit{IndianSubContinent}), 
    \item (6) Middle Eastern (\textit{Muslim}, \textit{Jewish}),
    \item (7) Southern European (\textit{Hispanic}, \textit{Italian}),
    \item (8) Western \& Northern European (\textit{French}, \textit{Germanic}, \textit{Nordic}).
\end{itemize}

Note that Chinese ethnicity (defined in \textit{Ethnea}) is by default incorporated into the \textit{EastAsian} ethnicity in the Wikipedia data.
We further placed the 8 categories into 4 groups for journalists' ethnicity due to insufficient instances of identified journalists in news stories: 
(1) Asian (East Asian, Indian),
(2) Anglo (British),
(3) European (Eastern European, Southern European, Western \& Northern European),
(4) Other Unknown (African, Middle Eastern, Unknown).
We fit the specification of Model 5 with Anglo and Man used as the reference categories.\\


\subsubsection{Race in U.S. Census Data}

Similarly, we coded the race for authors and journalists using races defined in the 2010 U.S. Census data based on the last name \cite{ambekar2009name,sood2018predicting}. 
The four race categories: (1) Asian (\textit{api}; [note that \textit{api} denotes Asian and Pacific Islander]), (2) Black (\textit{black}), (3) Hispanic (\textit{hispanic}), (4) White (\textit{white}), are directly used to fit the specification of Model 5 with White and Man used as the reference categories. 

\begin{figure}[ht!] 
\centering
\includegraphics[trim=0mm 0mm 0mm 0mm, width=0.8\linewidth]{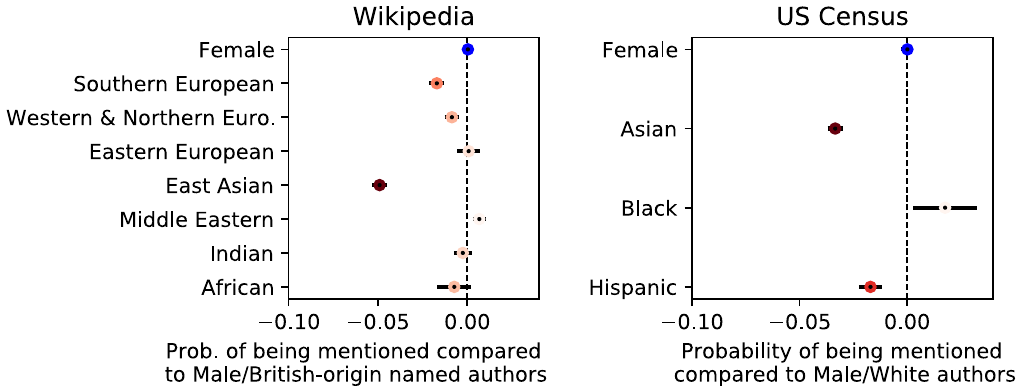}
\caption{Average marginal effects in mention probability for ethnicity and race, using Wikipedia data for coding ethnicity (\textbf{Left}) or U.S. Census data for coding race (\textbf{Right}). The analysis is based on a Model 5 fitted to all 524,052 observations. Note that gender is still inferred using \textit{Ethnea}. There are less than 1,000 authors with names predicted to be Black (Table~\ref{tab:census-Black}), which limits our ability to detect meaningful differences for this racial group.}
\label{fig:validation}
\end{figure}

Fig.~\ref{fig:validation} shows the average marginal effects in mention rates for scholars with names having minority ethnicity (or race) compared to Anglo (or White) named authors. As neither tool infers gender, we thus report the result for gender here using \textit{Ethnea}'s labels.
Like the case of \textit{Ethnea}, we find strong evidence of disparities for Asian-associated names in author mentions in science news, highlighting the robustness of our findings in the main text.

\section{Supplementary Tables}

\begin{table}[ht!]
\centering
\begin{tabular}{|l|l|}
\hline
\textbf{Broad Ethnic Category}    & \textbf{Individual Ethnicity}          \\ \hline
African                  & \textit{African}                                                   \\ 
Anglo           & \textit{English}                                                   \\ 
Chinese                  & \textit{Chinese}                                                   \\ 
non-Chinese East Asian   & \textit{Indonesian}, \textit{Japanese}, \textit{Korean}, \textit{Mongolian}, \textit{Thai}, \textit{Vietnamese} \\
Eastern European         & \textit{Hungarian}, \textit{Romanian}, \textit{Slav} \\ 
Indian                   & \textit{Indian}                                                    \\ 
Middle Eastern           & \textit{Arab}, \textit{Israeli}, \textit{Turkish} \\ 
Southern European &     \textit{Hispanic}, \textit{Italian}, \textit{Greek} \\ 
Western \& Northern European & \textit{Baltic}, \textit{Dutch}, \textit{French}, \textit{German}, \textit{Nordic}   \\ 
Caribbean         & \textit{Caribbean} \\ 
Polynesian         & \textit{Polynesian} \\ 
Unknown                  & Note: names are unrecognized by \textit{Ethnea}.                   \\ \hline
\end{tabular}
\caption{26 individual ethnicities were grouped into 11 broad ethnic categories. The last two groups, Caribbean and Polynesian, were excluded due to less than 100 observations.}
\label{tab:ethnea-broad}
\end{table}

\begin{table}[ht!]
\centering
\begin{tabular}{|l|r|r|}
\hline
\textbf{Author Broad Ethnicity Category} & \textbf{\# Paper Authorships} & \textbf{\# Triplets} \\ \hline
Anglo          &  81,226 & 234,510 \\
Western \& Northern European & 39,007 & 106,331 \\
Southern European  & 19,109 & 51,134 \\
Chinese       &  16,054 & 43,039 \\
Middle Eastern      & 9,185 & 26,082  \\
Indian         & 7,505 & 21,314 \\
non-Chinese East Asian & 7,816  &  19,068 \\
Eastern European    & 6,315  & 17,251  \\
African         & 1,079 & 2,774 \\
Unknown Ethnicity        & 898 & 2,549 \\ \hline
Total           & 188,194  & 524,052 \\ \hline
\end{tabular}
\caption{The number of paper authorships and the total number of (story, paper, author) triplets for the 9 high-level ethnic groups. Note that there are 100,486 unique papers, with some counted twice or more for authorships. For example, if a paper has 3 authors and gets covered by 2 news stories, it contributes 3 (paper, author) pairs, and 6 (story, paper, author) triplets.}
\label{tab:ethnea-broad-paper-mention}
\end{table}

\begin{table}[ht!]
\centering
\begin{tabular}{|l|r|}
\hline
\textbf{Journalist Broad Ethnicity Category} & \textbf{\# Triplets} \\ \hline
Anglo           &       68,652 \\ 
Western \& Northern European    &  13,790 \\
Southern European    &        10,594 \\
Middle Eastern      &        3,494 \\
Eastern European    &       2,924 \\
Chinese     &              2,449 \\
Indian       &             2,409 \\
non-Chinese East Asian   &    910 \\
African         &           643 \\
Unknown Ethnicity        & 418,187 \\ \hline
Total     & 524,052 \\ \hline
\end{tabular}
\caption{The number of (story, paper, author) triplets in our regression data by journalists' ethnicity.}
\label{tab:ethnea-broad-journalist}
\end{table}

\begin{table}[ht!]
\centering
\begin{tabular}{|l|r|r|r|r|}
\hline
\textbf{Outlet Type} & \textbf{\# Outlets} & \textbf{Example Outlet} & \textbf{\# Triplets} & \textbf{Perc. Aut. Ment.} \\ \hline
Press Releases   & 21 & EurekAlert!  & 165,343 & 63.5\% \\
Science \& Technology  & 86 & MIT Technology Rev.  & 137,851 & 41.9\% \\
General News   &  181 & The New York Times & 220,858 & 24.2\% \\ \hline
\end{tabular}
\caption{The number of outlets, the number of (story, paper, author) triplets, and the percentage of triplets that have mentioned the author, for three outlet types. The full list of 288 outlets are available in Appendix \tref{tab:outlet-list}.}
\label{tab:outlet-type-avg-rate}
\end{table}



\begin{table*}[t] \centering 


\resizebox{1.00\textwidth}{!}{
\begin{tabular}{ 
l @{\extracolsep{3pt}} l 
S[table-align-text-post=false,table-format=2.3]
S[table-align-text-post=false,table-format=2.5]
S[table-align-text-post=false,table-format=2.5]
S[table-align-text-post=false,table-format=2.5]
S[table-align-text-post=false,table-format=2.5]
} 

&& {\textbf{Model 1}} & {\textbf{Model 2}} & {\textbf{Model 3}} & {\textbf{Model 4}} & {\textbf{Model 5}}\\ 
\cline{2-7} \\[-1.8ex]
 
\parbox[t]{0mm}{\multirow{11}{*}{\rotatebox[origin=c]{90}{{ \small \textsc{Author Demog.}}}}}

  & \cellcolor{blue!25}African & -0.457\sym{***} & -0.394\sym{***} & -0.388\sym{***} & -0.371\sym{***} & -0.366\sym{***} \\ 
  & \cellcolor{blue!25}Chinese & 0.132\sym{***} & 0.099\sym{***} & -0.054\sym{***} & -0.254\sym{***} & -0.376\sym{***} \\ 
  & \cellcolor{blue!25}non-Chinese East Asian & 0.015 & 0.123\sym{***} & 0.037\sym{*} & -0.179\sym{***} & -0.272\sym{***} \\ 
  & \cellcolor{blue!25}Eastern European & 0.211\sym{***} & 0.253\sym{***} & 0.149\sym{***} & -0.021 & -0.009 \\ 
  & \cellcolor{blue!25}Indian & 0.138\sym{***} & 0.154\sym{***} & 0.048\sym{**} & -0.020 & -0.011 \\ 
  & \cellcolor{blue!25}Middle Eastern & 0.100\sym{***} & 0.134\sym{***} & 0.083\sym{***} & 0.014 & 0.016 \\ 
  & \cellcolor{blue!25}Southern European & -0.003 & 0.041\sym{***} & -0.017 & -0.114\sym{***} & -0.138\sym{***} \\ 
  & \cellcolor{blue!25}Western \& Northern European & -0.002 & 0.070\sym{***} & 0.027\sym{**} & -0.047\sym{***} & -0.072\sym{***} \\ 
  & \cellcolor{blue!25}Unknown Ethnicity & -0.210\sym{***} & -0.275\sym{***} & -0.354\sym{***} & -0.380\sym{***} & -0.227\sym{***} \\
  & \cellcolor{blue!35}Female & -0.150\sym{***} & -0.188\sym{***} & -0.128\sym{***} & -0.012 & 0.003 \\ 
  & \cellcolor{blue!35}Unknown Gender & -0.188\sym{***} & -0.158\sym{***} & -0.128\sym{***} & -0.117\sym{***} & -0.113\sym{***} \\ 
 \cline{2-7} \vspace{-3mm} \\
  & Author rank &  & 0.00002\sym{***} & -0.00003\sym{***} & -0.00005\sym{***} & -0.0001\sym{***} \\ 
  & Affiliation rank &  & -0.0001\sym{***} & -0.0001\sym{***} & -0.0001\sym{***} & -0.00004\sym{***} \\ 
  & Affiliation international (location) &  & -0.271\sym{***} & -0.263\sym{***} & -0.300\sym{***} & -0.307\sym{***} \\ 
  & Affiliation unknown (location) &  & 0.072 & 0.046 & 0.026 & 0.056 \\ 
  & Not a top author &  & 0.176\sym{***} & 0.168\sym{***} & 0.031 & -0.090\sym{**} \\ 
  & Not a corresponding author &  & -1.116\sym{***} & -1.230\sym{***} & -1.255\sym{***} & -1.448\sym{***} \\ 
  & Corresponding status unknown &  & -1.250\sym{***} & -0.519\sym{***} & -0.445\sym{***} & -0.506\sym{***} \\ 
  & Last name length &  & -0.005\sym{***} & -0.007\sym{***} & -0.007\sym{***} & -0.010\sym{***} \\ 
  & Last name frequency &  & 0.004\sym{**} & 0.003 & 0.002 & 0.004\sym{*} \\ 
  & Is the paper solo authored? &  & -0.152\sym{***} & 0.508\sym{***} & 0.561\sym{***} & 0.683\sym{***} \\ 
  & First author position &  & 0.142\sym{***} & 0.295\sym{***} & 0.340\sym{***} & 0.397\sym{***} \\ 
  & Middle author position &  & -0.329\sym{***} & -0.451\sym{***} & -0.623\sym{***} & -0.814\sym{***} \\ 
  & Number of authors in the paper &  & -0.001\sym{***} & -0.004\sym{***} & -0.005\sym{***} & -0.007\sym{***} \\ 
 \cline{2-7} \vspace{-3mm} \\

\parbox[t]{0mm}{\multirow{5}{*}{\rotatebox[origin=c]{90}{{ \small \textsc{Jrn. Demog.}}}}}

  &\cellcolor{orange!25}Asian &  &  & -0.255\sym{***} & -0.250\sym{***} & -0.051 \\ 
  &\cellcolor{orange!25}European &  &  & -0.057\sym{**} & -0.010 & -0.033 \\ 
  &\cellcolor{orange!25}Other Unknown Ethnicity &  &  & 0.342\sym{***} & 0.327\sym{***} & 0.054\sym{*} \\ 
  &\cellcolor{orange!35}Female &  &  & -0.185\sym{***} & -0.126\sym{***} & -0.015 \\ 
  &\cellcolor{orange!35}Unknown Gender &  &  & 0.094\sym{***} & 0.118\sym{***} & 0.015 \\ 
 \cline{2-7} \vspace{-3mm} \\

  &\cellcolor{green!15}Year of news story (mention year) &  &  & -0.089\sym{***} & -0.088\sym{***} & -0.051\sym{***} \\ 
  &\cellcolor{green!15}Year gap between story and paper &  &  & -0.228\sym{***} & -0.216\sym{***} & -0.145\sym{***} \\ 
  &\cellcolor{green!15}Num. of papers mentioned in a story &  &  & -0.159\sym{***} & -0.153\sym{***} & -0.120\sym{***} \\
  &\cellcolor{green!15}News story length &  &  & 0.0001\sym{***} & 0.0001\sym{***} & 0.0002\sym{***} \\ 
  &\cellcolor{red!15}Flesch-Kincaid score &  &  & -0.0002\sym{***} & -0.001\sym{***} & -0.001\sym{***} \\ 
  &\cellcolor{red!15}Sentences per paragraph &  &  & -0.006\sym{***} & 0.011\sym{***} & 0.008\sym{***} \\ 
  &\cellcolor{red!15}Type-Token ratio &  &  & 0.864\sym{***} & 0.347\sym{***} & 0.300\sym{***} \\ 
 
 \vspace{-3.5mm} \\
 \cline{2-7} \vspace{-3mm} \\
 
  & \textit{Intercept} & -0.308\sym{***} & 0.448\sym{***} & 0.463\sym{***} & 1.319\sym{***} & 0.968\sym{***} \\ 
 
\cline{2-7} \\ [-1.8ex]

 {\hspace{3mm} }
 & {Fixed effects for paper keywords} & {No} & {No} & {No} & {\textit{Yes}} & {\textit{Yes}} \\
 & {Random effects for outlets and venues} & {No} & {No} & {No} & {No} & {\textit{Yes}}\\ 
  
\cline{2-7} \\
 [-1.8ex]
& Akaike Inf. Crit. & {709,086.7} & {664,229.8} & {580,589.5} & {565,155.4} & {511,537.0} \\ 
\end{tabular} 
}
\caption{Coefficients of five increasing-complexity regression models in predicting if the author is mentioned by name using 524,052 (story, paper, author) observations. All variables in Model 5, including 199 keywords, are provided in Appendix~\tref{tab:m5-reg-tab}. Significance levels: *** p$<$0.001, ** p$<$0.01, and * p$<$0.05.} 
\label{tab:main-result}
\end{table*}

\begin{table}[ht!]
\centering
\begin{tabular}{|l|r|r|r|}
\hline
\textbf{Author Name}    & \textit{\textbf{Ethnea}} & \textbf{U.S. Census} & \textbf{Wikipedia}  \\ \hline
Alana Lelo  &  African  &  White  &  Romance Language \\
Samuel Lawn  &  African  &  White  &  Anglo \\
Saka S Ajibola  &  African  &  Black  &  East Asian \\
Mosi Adesina Ifatunji  &  African  &  Black  &  African \\
Sebastian Giwa  &  African  &  White  &  African \\
Olabisi Oduwole  &  African  &  White  &  African \\
Chidi N. Obasi  &  African  &  White  &  African \\
Habauka M. Kwaambwa  &  African  &  Asian  &  African \\
Esther E Omaiye  &  African  &  White  &  African \\
Aurel T. Tankeu  &  African  &  White  &  Anglo \\ \hline
\end{tabular}
\caption{A random sample of 10 African-named authors predicted by \textit{Ethnea} (out of 908 in total in our data) and their ethnicity or race categories based on the U.S. census data or the Wikipedia data.}
\label{tab:ethnea-African}
\end{table}

\begin{table}[ht!]
\centering
\begin{tabular}{|l|r|r|r|}
\hline
\textbf{Author Name}  & \textbf{U.S. Census} & \textit{\textbf{Ethnea}} & \textbf{Wikipedia}  \\ \hline
E. Robinson  &  Black  &  Anglo  &  Anglo \\
Momar Ndao  &  Black  &  Southern European  &  African \\
Angela F Harris  &  Black  &  Anglo  &  Anglo \\
Daddy Mata-Mbemba  &  Black  &  Southern European  &  African \\
A Bolu Ajiboye  &  Black  &  African  &  African \\
Lasana T. Harris  &  Black  &  Anglo  &  Anglo \\
John M. Harris  &  Black  &  Anglo  &  Anglo \\
Edwin S Robinson  &  Black  &  Anglo  &  Anglo \\
Eric A. Coleman  &  Black  &  Anglo  &  Anglo \\
Mp Coleman  &  Black  &  Anglo  &  Anglo \\ \hline
\end{tabular}
\caption{A random sample of 10 Black authors predicted based on the U.S. census data (out of 892 in total in our data) and their ethnicity categories based on \textit{Ethnea} or the Wikipedia data.}
\label{tab:census-Black}
\end{table}

\begin{table}[ht!]
\centering
\begin{tabular}{|l|r|r|}
\hline
\textbf{Author Broad Ethnicity} & \textbf{After Filtering} & \textbf{Before Filtering} \\ \hline
Anglo                   & 0.447 & 0.442 \\
Western \& Northern European   & 0.203 & 0.205 \\
Southern European          & 0.098 & 0.104 \\
Chinese                  &  0.082 & 0.079  \\
Middle Eastern           &   0.050 & 0.049 \\
Indian                   &  0.041 & 0.041  \\
non-Chinese East Asian    &  0.036 & 0.037 \\
Eastern European           & 0.033 & 0.034 \\
African                    & 0.005 & 0.005  \\
Unknown Ethnicity           & 0.005 & 0.005 \\ \hline
\end{tabular}
\caption{The percentage of (story, paper, author) triplets for the 9 high-level ethnic groups, before and after data filtering. The two distributions are almost indistinguishable from each other based on a chi-square test (p = 1.0).}
\label{tab:before_after}
\end{table}

\clearpage

\section{Supplementary Figures}

\begin{figure}[ht!] 
\centering
\includegraphics[trim=0mm 0mm 0mm 0mm, width=\columnwidth]{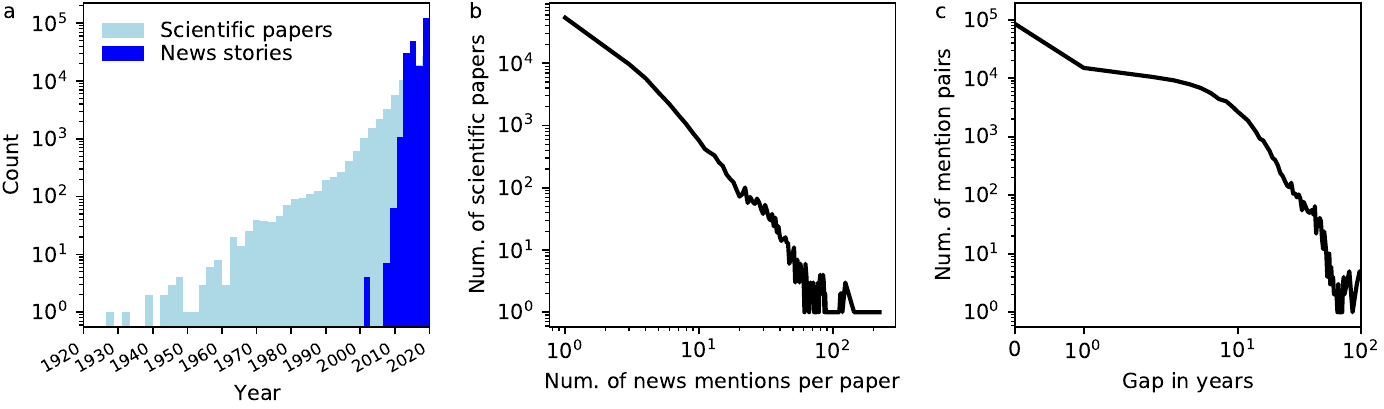}
\caption{\textbf{a,} The number of news stories and research papers in our mention date over time. \textbf{b,} The distribution of the number of news mentions per paper. \textbf{c,} The distribution of the \textit{year gap} between paper publication date and news story mention date for all 276,202 story-paper mention pairs in the final dataset.}
\label{fig:count}
\end{figure} 

\begin{figure*}[ht!] 
\centering
\includegraphics[trim=0mm 0mm 0mm 0mm, width=0.85\linewidth]{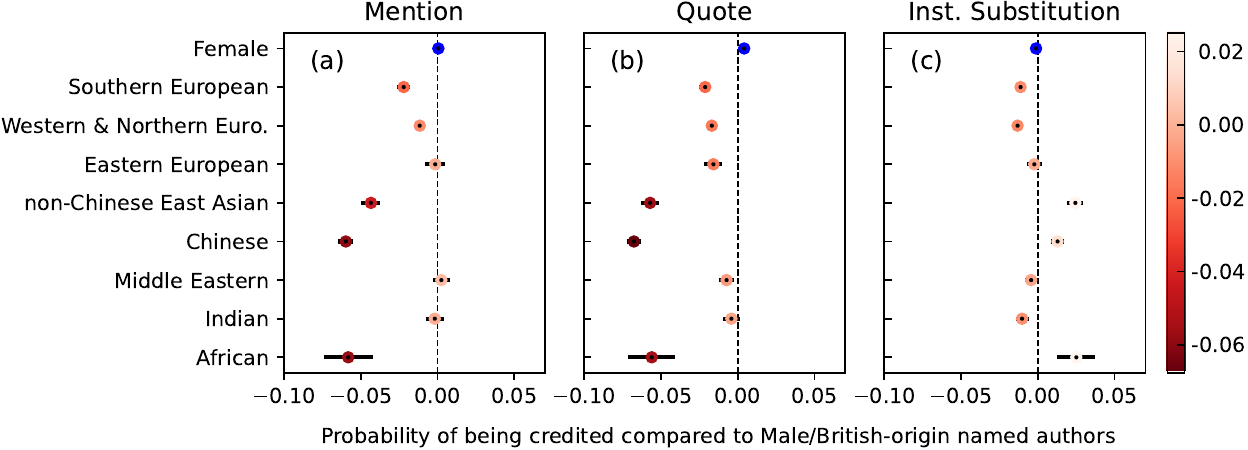}
\caption{The average marginal effects of ethnicity estimated based on 524,052 observations in the full data. 
Authors with minority-ethnicity names are less likely to be mentioned by name (\textbf{left}) or quoted (\textbf{middle}), and are more likely to be substituted by their institution (\textbf{right}).
A negative marginal effect indicates a decrease in probability compared to authors with Man (for gender) or Anglo (for ethnicity) names. The colors are proportional to the absolute probability changes.
\textit{Woman} is colored as blue to reflect its difference from ethnicity identities.
The error bars indicate 95\% bootstrapped confidence intervals.}
\label{fig:full-bias-result}
\end{figure*} 


\clearpage

\section{Supplementary Text}

\subsection{Associations of Control Variables with Author Mentions}

Although our focus is on ethnicity and gender, we find that many control variables are strongly associated with author mention rates. 
Examining the influence of these factors can lead to a better understanding of the mechanisms at play in science reporting.
Below we interpret their effects based on Model 5 (\tref{tab:main-result}) along three themes: (1) prestige related inequality, (2) impact of co-authorship, and (3) story content effects.

Not surprisingly, being designated as the corresponding author is positively associated with name mentions. 
Scholars who have a high professional rank or are affiliated with prestigious institutions receive outsized name mentions in science news when their research is covered. 
Popular authors whose research received many press coverage are more likely to be mentioned by name. 
This result suggests that the benefits of status, the so-called ``Matthew Effect'' \cite{merton1968matthew}, persist even after publication.

Having more co-authors on a paper has a negative effect on the author being mentioned.
Compared to the last author position, the first author is more likely to be mentioned by name, whereas the middle author is less likely to be named. 
The observed first position effect might due to the fact that, among papers (excluding solo-author papers) that have the corresponding author information, 59.9\% have the first author as corresponding and only 36.1\% have the last author as corresponding.
Solo-authored papers have been decreasing over time and are associated with lower impact on average \cite{greene2007demise,milojevic2014principles}.
However, our results highlight an underappreciated benefit---conditional on a paper being referenced in the news, a solo author is significantly more likely to be mentioned compared to authors of a multi-author paper.
Although seemingly counter to previous studies, it has a natural explanation---there is only one person to mention if need be.

The coefficients for story features point to the multifaceted nature of science reporting. 
Although the volume of science reporting is increasing over time (Fig.~\ref{fig:count}a), journalists tend to mention authors less frequently in later years.
At the same time, while older papers are still discussed in the media (Fig.~\ref{fig:count}c), journalists are less likely to mention authors of these studies as often. 
When more papers are referenced in a story, their authors are less likely to be mentioned. We hypothesize that such stories are often citing multiple scientific papers to construct a large narrative and thus those papers are only mentioned in passing. 
Longer stories are more likely to mention author names as they have more space to engage the authors.

\subsection{Does It Matter Who Is Reporting?}

Understanding whether ethnic disparities are related to journalists' own identities may help uncover the mechanisms producing them.
First, journalists of different ethnicities may differ in their overall tendencies to mention authors. If so, disparities may be driven by the composition of journalists. 
Our fullest model controls for journalists' name-inferred ethnicity, and shows that journalists with minority-identity associated names are not more or less likely to mention authors compared with journalists with Man or Anglo names (\tref{tab:main-result}, Model 5). 
We also note that, when dropping controls for outlets (Models 3-4), journalists' ethnicities become significant, suggesting that journalists' differential behavior might be explained by variations at the outlet level, \textit{i.e.} certain news outlets mention authors more or less often and certain groups of journalists are under- or over-represented in those outlets.

Second, there might exist interactive relationships between authors' and journalists' ethnic identities. 
One intuitive hypothesis, which we call ``ethnic hierarchy,'' is that all journalists, regardless of their perceived ethnicity, prefer to mention Anglo-named scholars over others. 
On the other hand, journalists may prefer to mention authors of the same ethnicity, which we call ``ethnic homophily''. Evidence for demographic homophily is pervasive\cite{mcpherson2001birds}. For example, concordance of gender identities between actors has been found to predict outcomes in domains such as healthcare\cite{greenwood2018patient}.
However, the relatively small number of cases of identified journalists (\tref{tab:ethnea-broad-journalist}) prevents us from including the full interactions between author's and journalist's ethnicities in the model.
The present study thus lacks the evidence to suggest either ethnic hierarchy or homophily hypotheses. However, this is an important avenue for future research. 

\bibliography{references}

\clearpage

\appendix

\section{Tables}

\begin{spacing}{1.0}
\begin{center}
\begin{longtable}{llr}
\caption{A random sample of 10 names for each of the 24 individual ethnicities and the ``Unknown'' category. All 6 MONGOLIAN names in our data are shown here.}
\label{tab:eth-names}\\
\textbf{Ethnicity} & \textbf{Name Example} & \textbf{Gender} \\ \hline
AFRICAN  &  Dora Wynchank  &  F \\
 &  Benjamin D. Charlton  &  M \\
 &  J. Nwando Olayiwola  &  unknown \\
 &  Ayodeji Olayemi  &  M \\
 &  Elizabeth Gathoni Kibaru  & F \\
 &  Christopher Changwe Nshimbi & M \\
 &  Naganna Chetty & unknown \\
 &  Benjamin Y. Ofori & M \\
 &  Khadijah Essackjee & F \\
 &  Jeanine L. Marnewick  &  F \\
 &  Habtamu Fekadu Gemede  &  M \\ \hline
ARAB  &  Zaid M. Abdelsattar  &  M \\
 &  Alireza Dirafzoon  &  M \\
 &  Ahmad Nasiri  &  M \\
 &  Saleh Aldasouqi  &  M \\
 &  Ibrahim A. Arif  &  M \\
 &  Sameer Ahmed  &  M \\
 &  A Elgalib  &  unknown \\
 &  Taha Adnan Jan  &  M \\
 &  Mohsen Taghizadeh  &  M \\
 &  Behnam Nabet  &  M \\ \hline
BALTIC  &  Skirmantas Kriaucionis  &  M \\
 &  Airidas Korolkovas  &  M \\
 &  Egle Cekanaviciute  &  F \\
 &  Arunas L. Radzvilavicius  &  M \\
 &  Ieva Tolmane  &  F \\
 &  Alberts B  &  M \\
 &  Gediminas Gaigalas  &  M \\
 &  Armandas Balcytis  &  unknown \\
 &  Ruta Ganceviciene  &  F \\
 &  Andrius Pašukonis  &  M \\ \hline
CHINESE  &  Chin Hong Tan  &  unknown \\
 &  Li Yuan  &  unknown \\
 &  Yalin Li  &  unknown \\
 &  Xian Adiconis  &  unknown \\
 &  Philip Sung-En Wang  &  M \\
 &  Xiaohui Ni  &  unknown \\
 &  Minghua Li  &  unknown \\
 &  Fang Fang Zhang  &  F \\
 &  Li-Qiang Qin  &  M \\
 &  Jian Tan  &  unknown \\ \hline
DUTCH  &  Pieter A. Cohen  &  M \\
 &  I. Vandersmissen  &  unknown \\
 &  Marleen Temmerman  &  F \\
 &  Gerard 't Hooft  &  M \\
 &  A. Yool  &  unknown \\
 &  G. A W Rook  &  unknown \\
 &  Fatima Foflonker  &  F \\
 &  Mirjam Lukasse  &  F \\
 &  Sander Kooijman  &  M \\
 &  Izaak D. Neveln  &  M \\ \hline
ENGLISH  &  Isabel Hilton  &  F \\
 &  Gavin J. D. Smith  &  M \\
 &  Katherine A. Morse  &  F \\
 &  Andrew S. Bowman  &  M \\
 &  T. M. L. Wigley  &  unknown \\
 &  Francis Markham  &  M \\
 &  Neil T. Roach  &  M \\
 &  Brooke Catherine Aldrich  &  F \\
 &  Vaughn I. Rickert  &  M \\
 &  Kellie Morrissey  &  F \\ \hline
FRENCH  &  Lucas V. Joel  &  M \\
 &  Daniel Clery  &  M \\
 &  Pierre Jacquemot  &  M \\
 &  Scott Le Vine  &  M \\
 &  Nathalie Dereuddre-Bosquet  &  F \\
 &  Stéphane Colliac  &  unknown \\
 &  Adelaide Haas  &  F \\
 &  Julie M. D. Paye  &  F \\
 &  Justine Lebeau  &  F \\
 &  Arnaud Chiolero  &  M \\ \hline
GERMAN  &  Laure Schnabel  &  F \\
 &  Jeff M. Kretschmar  &  M \\
 &  E. Homeyer  &  unknown \\
 &  Maren N. Vitousek  &  F \\
 &  D. Wild  &  unknown \\
 &  Hany K. M. Dweck  &  M \\
 &  E. M. Fischer  &  unknown \\
 &  Paul Marek  &  M \\
 &  Hans-Jörg Rheinberger  &  M \\
 &  Daniel James Cziczo  &  M \\ \hline
GREEK  &  Mary J. Scourboutakos  &  F \\
 &  Anita P Courcoulas  &  F \\
 &  Elgidius B. Ichumbaki  &  unknown \\
 &  Stavros G. Drakos  &  M \\
 &  Nikolaos Konstantinides  &  M \\
 &  Constantine Sedikides  &  M \\
 &  Maria A. Spyrou  &  F \\
 &  Panos Athanasopoulos  &  M \\
 &  Aristeidis Theotokis  &  M \\
 &  Amy H. Mezulis  &  F \\ \hline
HISPANIC  &  Mirela Donato Gianeti  &  F \\
 &  Julio Cesar de Souza  &  M \\
 &  Paulina Gomez-Rubio  &  F \\
 &  José A. Pons  &  M \\
 &  Arnau Domenech  &  M \\
 &  Nicole Martinez-Martin  &  F \\
 &  Mauricio Arcos-Burgos  &  M \\
 &  Raquel Muñoz-Miralles  &  F \\
 &  Annmarie Cano  &  F \\
 &  Merika Treants Koday  &  F \\ \hline
HUNGARIAN  &  Andrea Tabi  &  F \\
 &  Róbert Erdélyi  &  M \\
 &  Gabor G. Kovacs  &  M \\
 &  Xenia Gonda  &  F \\
 &  Erzsébet Bukodi  &  unknown \\
 &  Julianna M. Nemeth  &  F \\
 &  Ian K. Toth  &  M \\
 &  Zoltan Arany  &  M \\
 &  Cory A. Toth  &  M \\
 &  Ashley N. Bucsek  &  unknown \\ \hline
INDIAN  &  Sachin M. Shinde  &  M \\
 &  Govindsamy Vediyappan  &  M \\
 &  Ashish K. Jha  &  M \\
 &  Tamir Chandra  &  M \\
 &  Hariharan K. Iyer  &  M \\
 &  Chanpreet Singh  &  unknown \\
 &  Ravi Chinta  &  M \\
 &  Madhukar Pai  &  M \\
 &  Lalitha Nayak  &  F \\
 &  Ravi Dhingra  &  M \\ \hline
INDONESIAN  &  Dewi Candraningrum  &  unknown \\
 &  Richard Tjahjono  &  M \\
 &  T. A. Hartanto  &  unknown \\
 &  Johny Setiawan  &  M \\
 &  Truly Santika  &  unknown \\
 &  Chairul A. Nidom  &  unknown \\
 &  Christine Tedijanto  &  F \\
 &  Alberto Purwada  &  M \\
 &  Ardian S. Wibowo  &  M \\
 &  Anna I Corwin  &  F \\ \hline
ISRAELI  &  Ron Lifshitz  &  M \\
 &  Martin H. Teicher  &  M \\
 &  Ruth H Zadik  &  F \\
 &  Gil Yosipovitch  &  M \\
 &  Mor N. Lurie-Weinberger  &  unknown \\
 &  J. Tarchitzky  &  unknown \\
 &  Ilana N. Ackerman  &  F \\
 &  B. Trakhtenbrot  &  unknown \\
 &  Yoram Barak  &  M \\
 &  Mendel Friedman  &  M \\ \hline
ITALIAN  &  Tiziana Moriconi  &  F \\
 &  Marco Gobbi  &  M \\
 &  Marco De Cecco  &  M \\
 &  F. Govoni  &  unknown \\
 &  Theodore L. Caputi  &  M \\
 &  Mark A Bellis  &  M \\
 &  Fernando Migliaccio  &  M \\
 &  Julien Granata  &  M \\
 &  Jennifer M. Poti  &  F \\
 &  Brendan Curti  &  M \\ \hline
JAPANESE  &  Takuji Yoshimura  &  M \\
 &  Maki Inoue-Choi  &  F \\
 &  Masaaki Sadakiyo  &  M \\
 &  Moeko Noguchi-Shinohara  &  F \\
 &  Naoto Muraoka  &  M \\
 &  Shigeki Kawai  &  M \\
 &  Koji Mikami  &  M \\
 &  Masayoshi Tokita  &  M \\
 &  Naohiko Kuno  &  M \\
 &  Saba W. Masho  &  F \\ \hline
KOREAN  &  Jih-Un Kim  &  M \\
 &  Hanseon Cho  &  unknown \\
 &  Hyung-Soo Kim  &  M \\
 &  Yun-Hee Youm  &  F \\
 &  Yoon-Mi Lee  &  unknown \\
 &  Soo Bin Park  &  F \\
 &  Yungi Kim  &  unknown \\
 &  Woo Jae Myung  &  unknown \\
 &  Kunwoo Lee  &  unknown \\
 &  Sandra Soo-Jin Lee  &  F \\ \hline
MONGOLIAN  &  C. Jamsranjav  &  unknown \\
 &  Jigjidsurengiin Batbaatar  &  unknown \\
 &  Khishigjav Tsogtbaatar  &  unknown \\
 &  Migeddorj Batchimeg  &  unknown \\
 &  Tsolmon Baatarzorig  &  unknown \\
NORDIC  &  Steven G. Rogelberg  &  M \\
 &  Kirsten K. Hanson  &  F \\
 &  Jan L. Lyche  &  M \\
 &  Morten Hesse  &  M \\
 &  Karolina A. Aberg  &  F \\
 &  Britt Reuter Morthorst  &  F \\
 &  Kirsten F. Thompson  &  F \\
 &  Shelly J. Lundberg  &  F \\
 &  G Marckmann  &  unknown \\
 &  David Hägg  &  M \\ \hline
ROMANIAN  &  Afrodita Marcu  &  F \\
 &  Iulia T. Simion  &  F \\
 &  Liviu Giosan  &  M \\
 &  Alina Sorescu  &  F \\
 &  Liviu Giosan  &  M \\
 &  Mircea Ivan  &  M \\
 &  Dana Dabelea  &  F \\
 &  Constantin Rezlescu  &  M \\
 &  Christine A. Conelea  &  F \\
 &  R. A. Popescu  &  unknown \\ \hline
SLAV  &  Noémi Koczka  &  F \\
 &  Mikhail G Kolonin  &  M \\
 &  Richard Karban  &  M \\
 &  Branislav Dragović  &  M \\
 &  H Illnerová  &  unknown \\
 &  Marte Bjørk  &  F \\
 &  Jacek Niesterowicz  &  M \\
 &  Justin R. Grubich  &  M \\
 &  Mikhail Salama Hend  &  M \\
 &  Snejana Grozeva  &  F \\ \hline
THAI  &  Piyamas Kanokwongnuwut  &  unknown \\
 &  Clifton Makate  &  M \\
 &  Noppol Kobmoo  &  unknown \\
 &  Kabkaew L. Sukontason  &  unknown \\
 &  Aroonsiri Sangarlangkarn  &  unknown \\
 &  Yossawan Boriboonthana  &  unknown \\
 &  Ekalak Sitthipornvorakul  &  unknown \\
 &  Tony Rianprakaisang  &  M \\
 &  Apiradee Honglawan  &  F \\
 &  Wonngarm Kittanamongkolchai  &  unknown \\ \hline
TURKISH  &  Iris Z. Uras  &  F \\
 &  Metin Gurcan &  unknown \\
 &  Mustafa Sahmaran  &  M \\
 &  Pinar Akman  &  F \\
 &  Joshua Aslan  &  M \\
 &  Selin Kesebir  &  F \\
 &  Tan Yigitcanlar  &  unknown \\
 &  Thembela Kepe  &  unknown \\
 &  Ulrich Rosar  &  M \\
 &  Selvi C. Ersoy  &  F \\ \hline
VIETNAMESE  &  Huong T. T. Ha  &  unknown \\
 &  Vu Van Dung  &  M \\
 &  H ChuongKim  &  unknown \\
 &  Daniel W. Giang  &  M \\
 &  Nhung Thi Nguyen  &  unknown \\
 &  V. Phan  &  unknown \\
 &  Oanh Kieu Nguyen  &  F \\
 &  Phuc T. Ha  &  M \\
 &  Bich Tran  &  unknown \\
 &  Oanh Kieu Nguyen  &  F \\ \hline
Unknown  &  Gene Y. Fridman  &  M \\
 &  Judith Glück  &  F \\
 &  Noor Edi Widya Sukoco  &  unknown \\
 &  Charlene Laino  &  F \\
 &  Benoît Bérard  &  unknown \\
 &  David Zünd  &  M \\
 &  Katarzyna Adamala  &  F \\
 &  K.A. Godfrin  &  unknown \\
 &  Shadd Maruna  &  M \\
 &  Mariette DiChristina  &  F \\ \hline
\end{longtable}
\end{center}
\end{spacing}

\clearpage

\begin{spacing}{1.0}
\begin{center}
\begin{longtable}{@{\extracolsep{5pt}}lr}
\caption{The 288 U.S.-based outlets are grouped into 3 categories based on their topics of reports. Note that other 135 U.S.-based outlets, which are not shown in this table, are excluded in our analyses due to technical limitations in accessing sufficient volumes of their content (e.g., view-limited paywalls or anti-crawling mechanisms).}
\label{tab:outlet-list}\\
  
\textbf{Outlet} & \textbf{Type} \\ \hline
OnMedica & Sci. \& Tech. \\
Huffington Post & General News \\
KiiiTV 3 & General News \\
Carbon Brief & Sci. \& Tech. \\
PR Newswire & Press Releases \\
Nutra Ingredients USA & Sci. \& Tech. \\
The Bellingham Herald & General News \\
CNN News & General News \\
Health Medicinet & Press Releases \\
Herald Sun & General News \\
EurekAlert! & Press Releases \\
AJMC & Press Releases \\
The University Herald & General News \\
Lincoln Journal Star & General News \\
Cardiovascular Business & Sci. \& Tech. \\
MinnPost & General News \\
CNET & Sci. \& Tech. \\
Infection Control Today & Sci. \& Tech. \\
Science 2.0 & Sci. \& Tech. \\
Lexington Herald Leader & General News \\
Statesman.com & General News \\
Nanowerk & Press Releases \\
The San Diego Union-Tribune & General News \\
The Daily Beast & General News \\
Lab Manager & Press Releases \\
SDPB Radio & General News \\
New Hampshire Public Radio & General News \\
Health Day & Press Releases \\
Rocket News & General News \\
KPBS & General News \\
Technology.org & Press Releases \\
UPI.com & General News \\
WUWM & General News \\
Central Coast Public Radio & General News \\
The Hill & General News \\
The Epoch Times & General News \\
Biospace & Sci. \& Tech. \\
Minyanville: Finance & General News \\
Nature World News & Sci. \& Tech. \\
New York Post & General News \\
Action News Now & General News \\
WUNC & General News \\
Futurity & Press Releases \\
Reason & General News \\
azfamily.com & General News \\
Idaho Statements & General News \\
Google News & General News \\
Tri States Public Radio & General News \\
American Physical Society - Physics & Press Releases \\
KTEP El Paso & General News \\
LiveScience & Sci. \& Tech. \\
KUNC & General News \\
The Daily Meal & Sci. \& Tech. \\
AOL & General News \\
Women's Health & Sci. \& Tech. \\
Prevention & Sci. \& Tech. \\
ECN & Sci. \& Tech. \\
Iowa Public Radio & General News \\
Becker's Hospital Review & Sci. \& Tech. \\
7th Space Family Portal & Press Releases \\
Springfield News Sun & General News \\
Environmental News Network & Press Releases \\
Sky Nightly & Sci. \& Tech. \\
Quartz & Sci. \& Tech. \\
Benzinga & General News \\
Headlines \& Global News & General News \\
The Denver Post & General News \\
Science Daily & Press Releases \\
The Advocate & General News \\
ABC News & General News \\
Newswise & Press Releases \\
hellogiggles.com & General News \\
WLRN & General News \\
EarthSky & Sci. \& Tech. \\
Becker's Spine Review & Sci. \& Tech. \\
MIT News & Press Releases \\
MarketWatch & General News \\
Arstechnica & Sci. \& Tech. \\
Journalist's Resource & Sci. \& Tech. \\
Northern Public Radio & General News \\
Everyday Health & Sci. \& Tech. \\
Star Tribune & General News \\
TCTMD & Sci. \& Tech. \\
The Verge & General News \\
She Knows & General News \\
SeedQuest & Sci. \& Tech. \\
Tech Times & Sci. \& Tech. \\
Witchita's Public Radio & General News \\
Oncology Nurse Advisor & Sci. \& Tech. \\
Delmarva Public Radio & General News \\
Medical Daily & Sci. \& Tech. \\
Homeland Security News Wire & General News \\
Discover Magazine & Sci. \& Tech. \\
Washington Post & General News \\
MSN & General News \\
Hawaii News Now & General News \\
The Daily Caller & General News \\
News Tribune & General News \\
The Fresno Bee & General News \\
King 5 & General News \\
Star-Telegram & General News \\
CNBC & General News \\
Salon & General News \\
WJCT & General News \\
WVPE & General News \\
KTEN & General News \\
Wired.com & General News \\
Daily Kos & General News \\
USA Today & General News \\
Men's Health & Sci. \& Tech. \\
Boise State Public Radio & General News \\
Voice of America & General News \\
PR Web & Press Releases \\
Georgia Public Radio & General News \\
FiveThirtyEight & General News \\
Public Radio International & General News \\
Harvard Business Review & General News \\
Inverse & General News \\
Doctors Lounge & Sci. \& Tech. \\
North East Public Radio & General News \\
The Charlotte Observer & General News \\
National Geographic & Sci. \& Tech. \\
Pharmacy Times & Sci. \& Tech. \\
Popular Science & Sci. \& Tech. \\
ABC Action News WFTS Tampa Bay & General News \\
News Channel & General News \\
The University of New Orleans Public Radio & General News \\
Mic & General News \\
Health Canal & Sci. \& Tech. \\
KOSU & General News \\
Raleigh News and Observer & General News \\
The Atlantic & General News \\
newsmax.com & General News \\
Yahoo! Finance USA & General News \\
Government Executive & General News \\
International Business Times & General News \\
Emaxhealth.com & Press Releases \\
Newsweek & General News \\
FOX News & General News \\
The New York Observer & General News \\
Sign of the Times & General News \\
The Inquisitr & General News \\
ABC News 15 Arizona & General News \\
Parent Herald & General News \\
The ASCO Post & Sci. \& Tech. \\
Clinical Advisor & Sci. \& Tech. \\
Slate Magazine & General News \\
NPR & General News \\
Health & Sci. \& Tech. \\
Dayton Daily News & General News \\
Guardian Liberty Voice & General News \\
Belleville News-Democrat & General News \\
Yahoo! News & General News \\
WCBE & General News \\
Buzzfeed & General News \\
Sci-News & Sci. \& Tech. \\
The Seattle Times & General News \\
Philly.com & General News \\
Renal \& Urology News & Sci. \& Tech. \\
Arizona Public Radio & General News \\
Interlochen Public Radio & General News \\
12 News KBMT & General News \\
New York Magazine & General News \\
Medium US & General News \\
KPCC : Southern California Public Radio & General News \\
2 Minute Medicine & Sci. \& Tech. \\
Pediatric News & Sci. \& Tech. \\
redOrbit & Sci. \& Tech. \\
Insurance News Net & General News \\
Drug Discovery and Development & Sci. \& Tech. \\
USNews.com & General News \\
Yahoo! & General News \\
The Body & Sci. \& Tech. \\
GEN & Sci. \& Tech. \\
Pacific Standard & General News \\
Northwest Indiana Times & General News \\
Psychology Today & Sci. \& Tech. \\
Oregon Public Broadcasting & General News \\
Mother Nature Network & Sci. \& Tech. \\
Pressfrom & General News \\
Physician's Weekly & Sci. \& Tech. \\
Pettinga: Stock Market & General News \\
Winona Daily News & General News \\
Runner's World & Sci. \& Tech. \\
Bio-Medicine.org & Press Releases \\
Alternet & General News \\
Mother Jones & General News \\
The Wichita Eagle & General News \\
Cornell Chronicle & Press Releases \\
Politico Magazine & General News \\
Equities.com & General News \\
WBUR & General News \\
ABC 7 WKBW Buffalo & General News \\
Billings Gazette & General News \\
My Science & Sci. \& Tech. \\
The Week & General News \\
BioTech Gate & Sci. \& Tech. \\
Kansas City Star & General News \\
The Deseret News & General News \\
PBS & General News \\
Space.com & Sci. \& Tech. \\
Astrobiology Magazine & Sci. \& Tech. \\
Outside & General News \\
Value Walk & General News \\
WYPR & General News \\
Bustle & General News \\
Science World Report & Sci. \& Tech. \\
Inside Science & Sci. \& Tech. \\
Science Alert & Sci. \& Tech. \\
Breitbart News Network & General News \\
St. Louis Post-Dispatch & General News \\
HowStuffWorks & General News \\
Wyoming Public Radio & General News \\
UBM Medica & Sci. \& Tech. \\
Fight Aging! & Sci. \& Tech. \\
MIT Technology Review & Sci. \& Tech. \\
WVXU & General News \\
The Ecologist & Sci. \& Tech. \\
Alaska Despatch News & General News \\
Health Imaging & Sci. \& Tech. \\
Kansas City University Radio & General News \\
Christian Science Monitor & General News \\
Medicinenet & Sci. \& Tech. \\
WTOP & General News \\
Business Insider & General News \\
Real Clear Science & Sci. \& Tech. \\
Counsel \& Heal & Sci. \& Tech. \\
The Raw Story & General News \\
Medcity News & Sci. \& Tech. \\
Drugs.com & Sci. \& Tech. \\
Relief Web & Press Releases \\
SPIE Newsroom & Sci. \& Tech. \\
New York Daily News & General News \\
Newser & General News \\
The Sacramento Bee & General News \\
Vice & General News \\
R\&D & Sci. \& Tech. \\
KCENG12 & Sci. \& Tech. \\
Inc. & General News \\
Science/AAAS & Sci. \& Tech. \\
The Atlanta Journal Constitution & General News \\
Brookings & General News \\
Common Dreams & General News \\
Physician's Briefing & Press Releases \\
KERA News & General News \\
Space Daily & Sci. \& Tech. \\
Tech Xplore & Sci. \& Tech. \\
US News Health & Sci. \& Tech. \\
KUOW & General News \\
WRKF & General News \\
TIME Magazine & General News \\
Smithsonian Magazine & Sci. \& Tech. \\
Herald Tribune & General News \\
Lifehacker & General News \\
Fast Company & General News \\
Kansas Public Radio & General News \\
Omaha Public Radio & General News \\
New York Times & General News \\
Technology Networks & Sci. \& Tech. \\
Elite Daily & General News \\
Centre for Disease Research and Policy & Sci. \& Tech. \\
Business Wire & General News \\
KUNM & General News \\
CBS News & General News \\
Scientific American & Sci. \& Tech. \\
NBC News & General News \\
Sun Herald & General News \\
KRWG TV/FM & General News \\
TODAY & General News \\
Radio Acadie & General News \\
The Columbian & General News \\
Houston Chronicle & General News \\
WABE & General News \\
The Modesto Bee & General News \\
American Council on Science and Health & Sci. \& Tech. \\
WKAR & General News \\
Psych Central & Sci. \& Tech. \\
WebMD News & Sci. \& Tech. \\
Green Car Congress & Sci. \& Tech. \\
ABC News WMUR 9 & General News \\
Healthline & Sci. \& Tech. \\
Mongabay & Sci. \& Tech. \\
Vox.com & General News \\
WPTV 5 West Palm Beach & General News \\
Popular Mechanics & Sci. \& Tech. \\
PM 360 & Sci. \& Tech. \\
SFGate & General News \\
Seed Daily & Sci. \& Tech. \\ \hline
\end{longtable}
\end{center}
\end{spacing}

\clearpage

\begin{spacing}{1.0}
\begin{center}
\begin{longtable}{@{\extracolsep{10pt}}lSl}
  \caption{The coefficients of all variables (including 199 keywords) in Model 5 in predicting whether the author is mentioned by name in a news story referencing a research paper. Random effects for 288 outlets and 8,261 publication venues are also included in the model.}\\
  \label{tab:m5-reg-tab}\\
\\[-1.8ex]\hline
\hline \\[-1.8ex]
 & \multicolumn{2}{c}{\textit{Dependent variable:}} \\
\cline{2-3}
\\[-1.8ex] & \multicolumn{2}{c}{Is author mentioned} \\
\hline \\[-1.8ex]
  Author ethnicity African & -0.366 & p = 0.000 \\ 
  Author ethnicity Chinese & -0.376 & p = 0.000 \\ 
  Author ethnicity non-Chinese East Asian & -0.272 & p = 0.000 \\ 
  Author ethnicity Eastern European & -0.009 & p = 0.653 \\ 
  Author ethnicity Indian & -0.011 & p = 0.560 \\ 
  Author ethnicity Middle Eastern & 0.016 & p = 0.366 \\ 
  Author ethnicity Southern European & -0.138 & p = 0.000 \\ 
  Author ethnicity Western \& Northern Euro. & -0.072 & p = 0.000 \\ 
  Author ethnicity Unknown & -0.227 & p = 0.00002 \\ 
  Author gender Woman & 0.003 & p = 0.695 \\ 
  Author gender Unknown & -0.113 & p = 0.000 \\ 
  Reporter ethnicity Asian & -0.051 & p = 0.176 \\ 
  Reporter ethnicity European & -0.033 & p = 0.095 \\ 
  Reporter ethnicity Other Unknown & 0.054 & p = 0.047 \\ 
  Reporter gender Woman & -0.015 & p = 0.405 \\ 
  Reporter gender Unknown & 0.015 & p = 0.560 \\ 
  Last name length & -0.010 & p = 0.000 \\ 
  Last name frequency & 0.004 & p = 0.028 \\ 
  First author position & 0.397 & p = 0.000 \\ 
  Middle author position & -0.814 & p = 0.000 \\ 
  Is the paper solo authored & 0.683 & p = 0.000 \\ 
  Author rank & -0.0001 & p = 0.000 \\ 
  Not a top author & -0.090 & p = 0.004 \\ 
  Not a corresponding author & -1.448 & p = 0.000 \\ 
  Corresponding status unknown & -0.506 & p = 0.000 \\ 
  Affiliation rank & -0.00004 & p = 0.000 \\ 
  Affiliation international (location) & -0.307 & p = 0.000 \\ 
  Affiliation unknown (location) & 0.056 & p = 0.571 \\ 
  Number of authors in the paper & -0.007 & p = 0.000 \\ 
  Year of news story (mention year) & -0.051 & p = 0.000 \\ 
  Year gap between story and paper & -0.145 & p = 0.000 \\ 
  News story length & 0.0002 & p = 0.000 \\ 
  Num. of papers mentioned in a story & -0.120 & p = 0.000 \\ 
  Flesch-Kincaid score & -0.001 & p = 0.000 \\ 
  Sentences per paragraph & 0.008 & p = 0.00002 \\ 
  Type-Token ratio & 0.300 & p = 0.00000 \\ 
  Cell biology & 0.301 & p = 0.00000 \\ 
  Genetics & 0.001 & p = 0.980 \\ 
  Biology & 0.032 & p = 0.701 \\ 
  Body mass index & -0.329 & p = 0.00001 \\ 
  Health care & -0.183 & p = 0.0005 \\ 
  Disease & -0.103 & p = 0.018 \\ 
  Gerontology & -0.607 & p = 0.000 \\ 
  Population & -0.103 & p = 0.00003 \\ 
  Public health & -0.165 & p = 0.004 \\ 
  Medicine & -0.361 & p = 0.00001 \\ 
  Materials science & 0.352 & p = 0.001 \\ 
  Composite material & 0.162 & p = 0.188 \\ 
  Nanotechnology & 0.255 & p = 0.007 \\ 
  Cohort study & -0.009 & p = 0.861 \\ 
  Social psychology & -0.154 & p = 0.006 \\ 
  Cohort & 0.069 & p = 0.155 \\ 
  Psychological intervention & 0.009 & p = 0.879 \\ 
  Young adult & -0.309 & p = 0.00000 \\ 
  Family medicine & -0.306 & p = 0.00001 \\ 
  Cancer & -0.097 & p = 0.038 \\ 
  Surgery & -0.019 & p = 0.779 \\ 
  Randomized controlled trial & -0.095 & p = 0.062 \\ 
  Placebo & 0.019 & p = 0.790 \\ 
  Clinical trial & -0.105 & p = 0.190 \\ 
  Nursing & -0.288 & p = 0.002 \\ 
  Applied psychology & -0.425 & p = 0.011 \\ 
  Human factors and ergonomics & -0.220 & p = 0.061 \\ 
  Injury prevention & 0.335 & p = 0.002 \\ 
  Suicide prevention & 0.003 & p = 0.978 \\ 
  Psychiatry & -0.362 & p = 0.000 \\ 
  Occupational safety and health & -0.471 & p = 0.00002 \\ 
  Intensive care medicine & -0.286 & p = 0.001 \\ 
  Pediatrics & -0.241 & p = 0.0003 \\ 
  Hazard ratio & 0.266 & p = 0.00001 \\ 
  Confidence interval & -0.147 & p = 0.020 \\ 
  Retrospective cohort study & 0.148 & p = 0.039 \\ 
  Vaccination & 0.059 & p = 0.493 \\ 
  Psychology & 0.078 & p = 0.384 \\ 
  Perception & 0.185 & p = 0.021 \\ 
  Cognition & -0.117 & p = 0.034 \\ 
  Environmental health & -0.347 & p = 0.00000 \\ 
  Obesity & -0.203 & p = 0.003 \\ 
  Risk factor & 0.236 & p = 0.001 \\ 
  Quality of life & -0.035 & p = 0.702 \\ 
  Physical therapy & -0.094 & p = 0.095 \\ 
  Weight loss & -0.357 & p = 0.0001 \\ 
  Anatomy & 0.625 & p = 0.000 \\ 
  Mental health & 0.140 & p = 0.030 \\ 
  Psychosocial & 0.271 & p = 0.011 \\ 
  Anxiety & -0.334 & p = 0.00000 \\ 
  Distress & 0.269 & p = 0.012 \\ 
  Business & -0.660 & p = 0.00001 \\ 
  Public relations & -0.244 & p = 0.023 \\ 
  Marketing & 0.168 & p = 0.295 \\ 
  Immunology & -0.164 & p = 0.007 \\ 
  Global warming & -0.100 & p = 0.178 \\ 
  Economics & -0.040 & p = 0.741 \\ 
  Climatology & -0.254 & p = 0.003 \\ 
  Climate change & -0.461 & p = 0.000 \\ 
  General surgery & 0.008 & p = 0.960 \\ 
  Endocrinology & -0.154 & p = 0.007 \\ 
  Internal medicine & 0.341 & p = 0.000 \\ 
  Receptor & -0.160 & p = 0.055 \\ 
  Inflammation & 0.199 & p = 0.019 \\ 
  Stimulus  physiology  & 0.091 & p = 0.390 \\ 
  Immune system & 0.132 & p = 0.050 \\ 
  Meta analysis & -0.696 & p = 0.000 \\ 
  Sociology & 0.371 & p = 0.008 \\ 
  Gene & -0.131 & p = 0.031 \\ 
  Cancer research & -0.025 & p = 0.705 \\ 
  Breast cancer & 0.075 & p = 0.230 \\ 
  Cell & 0.385 & p = 0.00001 \\ 
  Diabetes mellitus & -0.062 & p = 0.159 \\ 
  Blood pressure & -0.127 & p = 0.177 \\ 
  Oncology & -0.172 & p = 0.049 \\ 
  Gynecology & -0.338 & p = 0.006 \\ 
  Communication & 0.319 & p = 0.006 \\ 
  Cognitive psychology & 0.002 & p = 0.983 \\ 
  Adverse effect & -0.092 & p = 0.208 \\ 
  Clinical endpoint & -0.626 & p = 0.000 \\ 
  Pharmacology & -0.392 & p = 0.0001 \\ 
  Virology & -0.330 & p = 0.0001 \\ 
  Risk assessment & 0.250 & p = 0.021 \\ 
  Transcription factor & 0.383 & p = 0.0001 \\ 
  Political science & -0.280 & p = 0.054 \\ 
  Ecology & 0.062 & p = 0.270 \\ 
  Geography & 0.018 & p = 0.864 \\ 
  Cross sectional study & -0.024 & p = 0.792 \\ 
  Odds ratio & -0.114 & p = 0.040 \\ 
  Comorbidity & -0.136 & p = 0.209 \\ 
  Environmental engineering & -0.452 & p = 0.005 \\ 
  Chemistry & 0.097 & p = 0.320 \\ 
  Medical emergency & -0.711 & p = 0.000 \\ 
  Physics & 0.131 & p = 0.214 \\ 
  Social science & 0.448 & p = 0.008 \\ 
  Ethnic group & 0.018 & p = 0.848 \\ 
  Labour economics & 0.380 & p = 0.015 \\ 
  Antibody & 0.274 & p = 0.008 \\ 
  Geomorphology & -0.160 & p = 0.102 \\ 
  Geophysics & 0.081 & p = 0.461 \\ 
  Geology & -0.312 & p = 0.002 \\ 
  Ranging & -0.113 & p = 0.215 \\ 
  Stroke & -0.003 & p = 0.974 \\ 
  Environmental resource management & -0.132 & p = 0.203 \\ 
  Type 2 diabetes & 0.169 & p = 0.053 \\ 
  Cardiology & 0.066 & p = 0.502 \\ 
  Molecular biology & 0.169 & p = 0.007 \\ 
  Developmental psychology & -0.043 & p = 0.499 \\ 
  Agriculture & -0.393 & p = 0.00002 \\ 
  Signal transduction & -0.188 & p = 0.053 \\ 
  Optoelectronics & -0.047 & p = 0.651 \\ 
  Psychotherapist & -0.413 & p = 0.004 \\ 
  Affect  psychology  & -0.319 & p = 0.003 \\ 
  Clinical psychology & -0.036 & p = 0.622 \\ 
  Anesthesia & -0.311 & p = 0.001 \\ 
  Atmospheric sciences & -0.029 & p = 0.774 \\ 
  In vivo & -0.117 & p = 0.192 \\ 
  Biochemistry & 0.0001 & p = 0.999 \\ 
  Analytical chemistry & -0.078 & p = 0.553 \\ 
  Neuroscience & 0.310 & p = 0.00001 \\ 
  Botany & -0.292 & p = 0.015 \\ 
  Gene expression & 0.242 & p = 0.017 \\ 
  Politics & 0.170 & p = 0.070 \\ 
  Demography & 0.339 & p = 0.000 \\ 
  Socioeconomic status & -0.345 & p = 0.00004 \\ 
  Mortality rate & -0.225 & p = 0.002 \\ 
  Virus & 0.066 & p = 0.494 \\ 
  Optics & 0.411 & p = 0.0004 \\ 
  Condensed matter physics & -0.591 & p = 0.000 \\ 
  Bioinformatics & -0.510 & p = 0.00001 \\ 
  Law & -0.111 & p = 0.494 \\ 
  Physical medicine and rehabilitation & -0.086 & p = 0.583 \\ 
  Stem cell & -0.056 & p = 0.496 \\ 
  Biodiversity & -0.167 & p = 0.022 \\ 
  Astrophysics & -1.033 & p = 0.000 \\ 
  Astronomy & -0.203 & p = 0.041 \\ 
  Radiology & -0.400 & p = 0.007 \\ 
  Pathology & -0.014 & p = 0.858 \\ 
  Proportional hazards model & -0.137 & p = 0.108 \\ 
  Chemotherapy & -0.662 & p = 0.00000 \\ 
  Predation & -0.196 & p = 0.029 \\ 
  Food science & -0.300 & p = 0.034 \\ 
  Artificial intelligence & 1.100 & p = 0.00002 \\ 
  Overweight & -0.049 & p = 0.571 \\ 
  Antibiotics & -0.043 & p = 0.710 \\ 
  Microbiology & 0.143 & p = 0.173 \\ 
  Zoology & 0.280 & p = 0.002 \\ 
  Paleontology & 0.200 & p = 0.016 \\ 
  Habitat & 0.546 & p = 0.000 \\ 
  Public administration & 0.924 & p = 0.00001 \\ 
  Ecosystem & -0.062 & p = 0.424 \\ 
  Economic growth & 0.095 & p = 0.450 \\ 
  Organic chemistry & 0.254 & p = 0.100 \\ 
  Government & -0.135 & p = 0.199 \\ 
  Autism & -0.140 & p = 0.133 \\ 
  Transplantation & 0.250 & p = 0.003 \\ 
  Gastroenterology & -0.297 & p = 0.022 \\ 
  Insulin & 0.018 & p = 0.849 \\ 
  Engineering & -0.268 & p = 0.133 \\ 
  Computer science & 0.072 & p = 0.529 \\ 
  Observational study & -0.154 & p = 0.111 \\ 
  Heart disease & 0.021 & p = 0.836 \\ 
  Epidemiology & -0.106 & p = 0.104 \\ 
  Obstetrics & 0.158 & p = 0.133 \\ 
  Pregnancy & -0.140 & p = 0.040 \\ 
  Fishery & 0.026 & p = 0.839 \\ 
  Alternative medicine & -0.243 & p = 0.041 \\ 
  Logistic regression & 0.385 & p = 0.00003 \\ 
  Offspring & 0.196 & p = 0.031 \\ 
  Mood & -0.287 & p = 0.002 \\ 
  Bacteria & 0.127 & p = 0.248 \\ 
  Prostate cancer & -0.400 & p = 0.00004 \\ 
  Evolutionary biology & 0.130 & p = 0.114 \\ 
  Phenomenon & 0.022 & p = 0.821 \\ 
  Longitudinal study & 0.027 & p = 0.758 \\ 
  Genome & 0.088 & p = 0.191 \\ 
  Mutation & 0.204 & p = 0.012 \\ 
  Pedagogy & -0.283 & p = 0.101 \\ 
  Dementia & -0.186 & p = 0.046 \\ 
  Relative risk & 0.121 & p = 0.109 \\ 
  Microeconomics & 0.536 & p = 0.003 \\ 
  Odds & 0.004 & p = 0.968 \\ 
  Feeling & 0.451 & p = 0.00004 \\ 
  Oceanography & -0.095 & p = 0.376 \\ 
  Emergency medicine & 0.029 & p = 0.759 \\ 
  Personality & -0.023 & p = 0.804 \\ 
  Prospective cohort study & -0.212 & p = 0.0003 \\ 
  Hippocampus & -0.046 & p = 0.650 \\ 
  Greenhouse gas & 0.006 & p = 0.948 \\ 
  Biomarker  medicine  & 0.409 & p = 0.00002 \\ 
  Myocardial infarction & -0.135 & p = 0.140 \\ 
  Socioeconomics & 0.297 & p = 0.015 \\ 
  Drug & 0.290 & p = 0.004 \\ 
  Environmental science & -0.368 & p = 0.0003 \\ 
  Epigenetics & -0.382 & p = 0.0002 \\ 
  Inorganic chemistry & -0.233 & p = 0.020 \\ 
  Emergency department & -0.205 & p = 0.028 \\ 
  Medical prescription & 0.270 & p = 0.002 \\ 
  Phenotype & 0.076 & p = 0.450 \\ 
  Constant & 0.968 & p = 0.000 \\ 
 \hline \\[-1.8ex] 
Observations & {524,052} \\ 
Log Likelihood & {-255,530.5} \\ 
Akaike Inf. Crit. & {511,537.0} \\ 
Bayesian Inf. Crit. & {514,195.3} \\ 
\hline
\hline \\[-1.8ex]
\end{longtable}
\end{center}
\end{spacing}